\documentclass[aps,prd,preprint,nofootinbib,longbibliography,superscriptaddress]{revtex4-1}  
\usepackage{graphicx}  
\usepackage{dcolumn}   
\usepackage{bm}        
\usepackage{amssymb}   
\usepackage{xcolor}
\usepackage[normalem]{ulem}
\usepackage{cancel}
\usepackage{amsmath}
\usepackage[english]{babel}
\usepackage{accents}
\usepackage{natbib}
\usepackage{hyperref}
\usepackage[bottom]{footmisc}
\usepackage[section]{placeins}

\newcommand{\be}{\begin{equation}}
\newcommand{\ee}{\end{equation}}

\newcommand{\e}{\mbox{e}}
\newcommand{\de}{\mbox{d}}

\newcommand{\spl}{\be\begin{split}}
\newcommand{\pa}{\partial}

\newcommand{\BlockRefA}{Penrose:1964wq,*Hawking:1966sx,*Hawking:1969sw,*Hawking:1973uf}
\newcommand{\BlockRefC}{Freidel:2005qe,*Oriti:2006se,*Oriti:2011jm,*Baratin:2011aa,*Krajewski:2012aw,*Kegeles:2015oua,*Kegeles:2016wfg}
\newcommand{\BlockRefE}{Thiemann:2007zz,*Rovelli:2004tv}
\newcommand{\BlockRefF}{Ooguri:1992eb,*Reisenberger:2000zc,*Freidel:2007py,*Krajewski:2010yq,*Baratin:2011tx,*Kaminski:2009fm,*Han:2011rf,*Baratin:2010wi,*Rovelli:2011eq,*Perez:2012wv}
\newcommand{\BlockRefH}{Gielen:2013kla,*Gielen:2013naa,*Gielen:2014uga}
\newcommand{\BlockRefJ}{Oriti:2016qtz,*Oriti:2016ueo}
\newcommand{\BlockRefK}{Bojowald:2001xe,*Ashtekar:2011ni,*Banerjee:2011qu,*Bojowald:2008zzb,*Bojowald:2012we}
\newcommand{\BlockRefL}{Konopka:2006hu}

\begin{document}

\begin{flushleft}
KCL-PH-TH/2017-41
\end{flushleft}

\date{\today}

\author{Marco de Cesare}
\email{marco.de\_cesare@kcl.ac.uk}
\affiliation{Department of Physics, King's College London, University
  of London, Strand WC2R 2LS, London, United Kingdom}
\author{Daniele Oriti}
\email{daniele.oriti@aei.mpg.de}
\affiliation{Max Planck Institute for Gravitational Physics (Albert Einstein Institute)\\ Am M\"uhlenberg 1, 14476 Potsdam, Germany, EU}
\author{Andreas~G.~A.~Pithis}
\email{andreas.pithis@kcl.ac.uk}
\affiliation{Department of Physics, King's College London, University
  of London, Strand WC2R 2LS, London, United Kingdom}
\affiliation{Max Planck Institute for Gravitational Physics (Albert Einstein Institute)\\ Am M\"uhlenberg 1, 14476 Potsdam, Germany, EU}
\affiliation{Perimeter Institute for Theoretical Physics, Waterloo, Ontario N2L 2Y5, Canada}
\author{Mairi Sakellariadou}
\email{mairi.sakellariadou@kcl.ac.uk}
\affiliation{Department of Physics, King's College London, University
  of London, Strand WC2R 2LS, London, United Kingdom}
  \affiliation{Perimeter Institute for Theoretical Physics, Waterloo, Ontario N2L 2Y5, Canada}

\title{Dynamics of anisotropies close to a cosmological bounce in quantum gravity} 

\begin{abstract}
We study the dynamics of perturbations representing deviations from perfect isotropy in the context of the emergent cosmology obtained from the group field theory formalism for quantum gravity. Working in the mean field approximation of the group field theory formulation of the Lorentzian EPRL model, we derive the equations of motion for such perturbations to first order. We then study these equations around a specific simple isotropic background, characterised by the fundamental representation of $\mbox{SU(2)}$, and in the regime of the effective cosmological dynamics corresponding to the bouncing region replacing the classical singularity, well approximated by the free GFT dynamics. In this particular example, we identify a region in the parameter space of the model such that perturbations can be large at the bounce but become negligible away from it, \emph{i.e.} when the background enters the non-linear regime. 
We also study the departures from perfect isotropy by introducing specific quantities, such as the \emph{surface-area-to-volume} ratio and the \emph{effective volume per quantum}, which make them quantitative.
\end{abstract}

\maketitle

\tableofcontents
\newpage
\section{Introduction}
Our current understanding of the early Universe based on standard cosmology is limited by the initial singularity, and more generally by the lack of control over the deep quantum gravity regime (around the Planck scale). The occurrence of spacetime singularities in general relativity (GR) is generic for matter satisfying suitable energy conditions, as shown by Hawking and Penrose \cite{\BlockRefA}. It is believed that quantum gravitational effects taking place at the Planck scale could lead to a resolution of the singularities, as first suggested in Ref.~\cite{DeWitt:1967aa}. In particular, this idea has been very fruitful in theoretical frameworks based on a fundamentally discrete quantum geometry, such as background independent approaches to quantum gravity. The first realisation was made within the context of loop quantum cosmology (LQC), a quantization inspired by loop quantum gravity (LQG) of the symmetry-reduced, cosmological sector of GR. The results obtained in such framework have shown that the initial singularity is replaced by a quantum bounce, which is a robust feature \cite{\BlockRefK}. This is a consequence of the existence of a ``gap'', \emph{i.e.} a non-vanishing minimum eigenvalue, in the spectrum of the geometric operators of LQG. Other models featuring a cosmological bounce, variously inspired by other approaches to quantum gravity, are reviewed in Ref.~\cite{Brandenberger:2016vhg} and references therein.

A particular background independent approach to quantum gravity in which a bouncing cosmological solution was found is group field theory (GFT). Remarkably, this result was obtained within the complete theory, \emph{i.e.} without symmetry reducing before quantising, as in LQC. Group field theory represents a higher dimensional generalisation of matrix models \cite{\BlockRefC}, like random tensor models, but further enriched by the group-theoretic data characterizing the quantum states of geometry also as in LQG. It can be understood indeed also as a second quantisation of LQG \cite{Oriti:2013aqa,\BlockRefE}. In this formulation, the elementary degrees of freedom (single particle states) are open spin network vertices, with their edges labelled by irreducible representations of a Lie group $G$ (typically $G=\mbox{SU(2)}$). They are dual to quantum tetrahedra, which can be understood as the basic building blocks, or quanta, of a spatial geometry.
GFTs for d-dimensional quantum gravity are field theories on a Lie group $G^d$, quantised e.g. by path integral methods. For a given GFT, the perturbative expansion of the path integral can be expressed as a sum of transition amplitudes of a given spin foam (SF) model, a covariant definition of the LQG dynamics \cite{\BlockRefF}.

At a fundamental level, spacetime is absent in the formulation of GFT, and rather an emergent concept, reminiscent of the way we understand collective phenomena in condensed matter physics. This general perspective has been advocated and outlined in Ref.~\cite{Oriti:2013jga}, and in related contexts in Ref.~\cite{\BlockRefL}. The programme of recovering cosmology from GFT along this perspective was started in Refs.~\cite{\BlockRefH,Sindoni:2014wya}. In following studies, Refs.~\cite{\BlockRefJ}, it was shown that GFT allows to derive an effective Friedmann equation from the evolution of the \emph{mean field}, within a generalised GFT formulation of the Lorentzian EPRL spin foam model (cf. Ref.~\cite{Engle:2007wy}). The mean field describes the collective dynamics of quanta of space/geometry. Solutions to the effective Friedmann equation describe then an emergent classical background obtained from a full theory of quantum gravity. As a general feature, the solutions of the model exhibit a quantum bounce which resolves the initial singularity, provided that the hydrodynamics approximation of GFT (where the mean field analysis is confined) holds \cite{\BlockRefJ,deCesare:2016axk,deCesare:2016rsf}. A number of other interesting results concerning this and other GFT models and their effective cosmological dynamics have been obtained; to mention a few: the acceleration phase after the bounce can be long lasting without the need to introduce an inflaton field, but purely driven by the GFT interactions \cite{deCesare:2016axk,deCesare:2016rsf}; the same interactions can produce a cyclic evolution for the universe, with infinite expansion and contraction phases \cite{deCesare:2016rsf}; the late time evolution of the universe can be captured entirely, in the mean field approximation, by simple condensates with wavefunctions which only excite a single spin \cite{Gielen:2016uft,Pithis:2016cxg}. On the other hand, so far, only perfectly homogeneous cosmologies have been considered, although progress to include inhomogeneities is underway \cite{Gielen:2014usa,Gielen:2015kua}. 

The cosmologies that have been considered so far are also, with the exception of the partial analysis in \cite{Pithis:2016cxg}, \lq isotropic\rq, thus governed by a single global degree of freedom, corresponding to the total volume (or scale factor) of the universe. The purpose of the present paper is to take a first step beyond this isotropic case, and study GFT perturbations involving anisotropic degrees of freedom around isotropic background configurations, focusing on the cosmological evolution around the bouncing region, and remaining within the mean field approximation of the full quantum dynamics. The dynamics of anisotropies is an extremely important issue in fundamental cosmology, and especially  in bouncing scenarios, where the growth of anisotropies close to the bounce is a problematic aspect if no mechanism exists to keep it under control. It is also a difficult one to tackle in full generality and in concrete physical terms in GFT condensate cosmology, for a number of reasons. Let us explain in slightly more detail what these difficulties are, and how we deal with them in this paper.

\

Homogeneity at the level of GFT condensate states corresponds to their defining property that the quantum data associated to the building blocks of space, i.e. quantized tetrahedra, thus their individual wavefunction, are the same for all of them. This is the simplest type of condensate state, before we consider depletion factors and perturbations, but also before we take into account entanglement between the same building blocks and therefore connectivity information. Such condensate states are characterized by a single collective wavefunction, whose domain of definition (defined in group-theoretic terms) is isomorphic to the minisuperspace of homogeneous continuum spatial geometries, in agreement with the cosmological interpretation of the states \cite{\BlockRefH}. The isotropic restriction is then a reduction of this domain of definition that results in a 1-dimensional manifold, parametrized by a single (group-theoretic) variable with the interpretation of (a function of) the volume or scale factor of a homogeneous and isotropic universe. This reduction can also be seen at the \lq microscopic\rq~level of the tetrahedra, as a geometric restriction. The specific isotropic restriction to be chosen is not unique, as it depends on a choice of map, instantiating the mentioned abstract isomorphism, between discrete data associated to each tetrahedron and the ones defining a metric at a point in a continuum space where this tetrahedron is imagined embedded, in turn depending on a choice of reference frame at the point (for details, see \cite{Gielen:2014ila} and the review \cite{Gielen:2016dss}). These ambiguities are not expected to produce different physical results, but they do suggest different definitions of isotropy. Two main ones have been used so far in the literature, one corresponding to GFT condensates made of tri-orthogonal tetrahedra in which three edges of the same length meet orthogonally at the same vertex \cite{Pithis:2016cxg}, and one to condensates made of equilateral tetrahedra \cite{\BlockRefJ}. This ambiguity is a first source of difficulty in the study of anisotropies in GFT condensate cosmology. 

A way to resolve this ambiguity, assuming it is physically significant, would be to define GFT observables with a clear and compelling geometric interpretation as capturing anisotropic degrees of freedom of continuum geometries, and extract their effective cosmological dynamics, in the same way in which it has been done in the isotropic case. However, this constriction has not yet been performed in this formalism, and it is the second, most crucial difficulty we have to face. As explained, all the degrees of freedom that are necessary to characterize anisotropies are, in principle, present in the condensate wavefunction; however, the intuition needed for the explicit construction of anisotropic geometric observables is hampered by the lack of any spacetime background structure and by the fact that, since GFT states involve generically a superposition of lattice structures (when they do encode such lattice structures via the connectivity information among the GFT quanta), they do not depend on any specific lattice (this has to be contrasted with the strategies followed in the related context of loop quantum cosmology \cite{Bojowald:2007ra}). For this reason, in this work we will often prefer the label \lq monochromatic\rq~over \lq isotropic\rq~and \lq non-monochromatic\rq~over \lq anisotropic\rq~when referring to the fundamental building blocks, in agreement with the definitions of Ref.~\cite{Brunnemann:aa}.
As a first step forward, we sidestep this difficulty by focusing on GFT perturbations of exact isotropic condensate wavefunctions. For any given definition of isotropy, anisotropies are going to be encoded by construction in such perturbations. A detailed study in terms of physical observables will be needed to explore how, exactly, but the fact itself is not in question. In particular, if one is only interested in the specific issue of whether anisotropies become dominant during the effective cosmological  evolution or remain subdominant compared to the isotropic background, it is enough to study the dynamics of such GFT perturbations and their relative amplitude compared to the background condensate wavefunction. This is the issue we focus on.
The other main difficulty we have to face (which affects the whole GFT cosmology programme, like all attempts to extract physics from GFT or spin foam models) is purely technical, and lies in the complication of the analytic expression of the interaction kernel for physically interesting (and thus more promising) models, here the EPRL model. The Lorentzian EPRL  vertex amplitude, in fact, has an analytic expression which can be variously expressed in integral form (where the integrals are over the Lorentz group and encode the covariant properties of the model) \cite{Speziale:2016axj}, but has not been put down explicitly as a function of its boundary data, which are usually given in terms of $\mbox{SU(2)}$ representations (to match the LQG form of quantum states). This is a serious limitation for the computation of transition amplitudes in the full theory, as well as for the solution of the classical dynamics of the corresponding GFT model, which would be our concern here. This is true both for the background dynamics and for that of the perturbations over it, so it is not sidestepped by our focus on the latter. Absent an explicit closed expression, the remaining option is to adopt a more phenomenological approach \cite{deCesare:2016axk,deCesare:2016rsf,Pithis:2016wzf,Pithis:2016cxg} and model the GFT interactions by simple functions that capture some of their essential features. Still, further results on the explicit evaluation of the fundamental dynamics will obviously be very important.
What allows us to make some progress on the study of cosmological anisotropies in this paper, despite of this technical difficulty, is another general fact about the effective cosmological evolution of GFT condensates. GFT interactions become subdominant, compared to the kinetic term, at low densities, thus for small values of the GFT field (or perturbations), beside for small values of the GFT coupling constants, of course. In turn, this is exactly what happens close to the cosmological bounce, i.e. for low enough values of the universe volume\footnote{This is true, of course, provided one remains within the hydrodynamic approximation, which is not so obvious, since low densities also means small average number of atoms in the condensate.}. This allows to study, as a first step, the dynamics and relative strength of anisotropic (non-monochromatic) perturbations on an isotropic (monochromatic) background close to the cosmological bounce, before the universe grows in size and \lq occupation number\rq~enough that the GFT interactions have to be included in the analysis. Luckily, as we mentioned, the behaviour of anisotropies close to the bounce, and their relative suppression compared to the isotropic background geometry away from it, where a Friedmann dynamics is expected to be a good description, is also the key question from the point of view of cosmological bouncing scenarios. The simple case we consider in this paper, therefore, is also of direct physical interest.

\

We briefly outline the plan of the paper. In Section~\ref{sec:1} we review the fundamentals of the GFT formalism needed for our applications, considering in particular the Lorentzian EPRL model based on $\mbox{SU(2)}$ states for quantum gravity \cite{Engle:2007wy} coupled to a massless scalar field \cite{Li:2017uao}, trying to be explicit about key technical details and assumptions, and then discuss how an effective cosmological dynamics can be extracted from it for simple GFT condensate states (using the scalar field as a relational clock), following \cite{\BlockRefJ,deCesare:2016axk,deCesare:2016rsf}. In particular, we discuss the symmetry assumptions on the condensate states and the isotropic reduction, leading to an effective (modified) Friedmann dynamics and a cosmic bounce. 
Section~\ref{sec:Perturbations} is where our new results are presented. We obtain the dynamics of non-monochromatic perturbations to first order. The dynamics is given by a system of four coupled linear differential equations. The terms arising from the linearisation of the interaction have non-constant coefficients depending on the background. Within this approximation, only one of the four faces of the tetrahedra can be perturbed. The others must match the spin of the background, due to a constraint given by the EPRL vertex. This is done in full generality (for EPRL-like models with a broad class of kinetic kernels). The dynamics can then be recast in a simpler and more compact form in the particular case of a local kinetic kernel and when considering a background with only spin $j=\frac{1}{2}$ being excited. Considering this specific case, in Section~\ref{sec:DynamicsAtBounce} we study the dynamics of non-monochromatic components at the bounce, where interactions are negligible. Hence, both the background and the perturbations satisfy linear equations of motion. In this regime, there is no need to impose the condition that perturbations are much smaller than the background. Thus, the perturbed geometry of the emergent spacetime can in principle be quite different from the one given by the background. An important result we obtained is the determination of a region of parameter space such that perturbations are bounded at all times while the background field grows unbounded. It is thus justified to neglect non-monochromaticity after the bounce, when the non-linear regime is entered. However, around the bounce the magnitude of the perturbations can be of the same order as the background, leading to interesting consequences. To illustrate this point, we compute geometric quantities such as the \emph{surface-area-to-volume ratio} and the \emph{effective volume per quantum}, which characterise the non-trivial corrections to the ``mean geometry'' of the elementary mono-chromatic constituents around the cosmological bounce. Our results do not depend on where we set the initial conditions for the non-monochromatic perturbations, \emph{i.e.} whether their amplitude is maximal at the bounce or in the contracting/expanding phase.
Finally, in the Conclusions we review our results and point at future directions. In the Appendices we give more mathematical details needed for our derivations.

\subsection*{Notation and conventions}
The GFT field for pure gravity is defined on four copies of a Lie group, to be extended when introducing matter degrees of freedom. Therefore, its argument is an array of four group elements $(g_1,g_2,g_3,g_4)$ which we will denote as $g_\nu$ for brevity. We use instead the notation $g_{i}$ for specific components of $g_\nu$. Analogously, we introduce the notations $j_\nu$, $m_\nu$, etc. for the spins and their counterparts with Latin indices. Sometimes more than four spins are considered. In that case, only Latin indices appear in the equations and refer to the individual spins.
$\hbar=G=c=1$ throughout the paper, unless otherwise stated.

\section{Brief review of the group field theory formalism and of its emergent cosmology}\label{sec:1}
\subsection{The GFT formalism}
In this Section, we review the basics of the GFT formalism for quantum gravity and only those aspects which will be needed for the cosmological applications considered in this work. The reader interested in a more comprehensive review of the mathematical and foundational aspects of GFT is referred to Refs.~\cite{Freidel:2005qe,Oriti:2006se,Krajewski:2012aw}. Reviews of the recent developments of the cosmological applications of GFT can be found in Refs.~\cite{Gielen:2016dss,Oriti:2016acw}.

The partition function of a GFT model with a complex scalar field $\varphi$ is
\be\label{eq:PathIntegral}
\mathcal{Z}=\int\mathcal{D}\varphi\mathcal{D}\overline{\varphi}\; \e^{-S[\varphi,\overline{\varphi}]},
\ee
where $S[\varphi,\overline{\varphi}]$ is the classical action and $\varphi$ is defined on $\mbox{SU(2)}^4 \times \mathbb{R}$
\be\label{eq:GFTfieldDef}
\varphi=\varphi(g_1,g_2,g_3,g_4;\phi) = \varphi(g_\nu;\phi)
\ee
$\phi$ is a matter field, which for simplicity can be taken so as to represent a massless scalar, used as a relational clock \cite{\BlockRefJ,Li:2017uao}. Just as for the discrete geometric data encoded in the group elements (and their conjugate variables), the interpretation of the real variable $\phi$ as a discretized matter field is grounded in the expression of the (Feynman) amplitudes of the model corresponding to the action $S$, which take the form of lattice gravity path integrals for gravity coupled to a massless scalar field. 
The field is invariant under the diagonal right action of $\mbox{SU(2)}$
\be\label{eq:right invariance}
\varphi(g_\nu h;\phi)=\varphi(g_\nu;\phi), \hspace{1em} \forall h\in \mbox{SU(2)}.
\ee
This \emph{right-invariance} property of the GFT field, and of the corresponding quantum states, can be understood as the usual gauge invariance on spin network states that characterizes any lattice gauge theory, and more geometrically as the closure of the tetrahedra dual to the vertices of the same spin network states (this becomes transparent in the formulation of the theory using non-commutative Lie algebra-valued flux variables \cite{Baratin:2011hp}). Mathematical details are contained in the Appendices~\ref{sec:Harmonic},~\ref{sec:Intertwiner}.

The classical GFT field and the wavefunctions associated to its quantum states are $L^2$ functions with respect to the Haar measure on the $\mbox{SU(2)}$ group manifold. Therefore they can be expanded, by Peter-Weyl theorem, in a basis of functions labeled by the irreducible representations of the same group (see Appendix~\ref{sec:Harmonic}). For quantum states, this is the decomposition in terms of spin network states. For the classical GFT field, this decomposition looks: 
\be\label{eq:GFTfield}
\varphi(g_\nu;\phi)=\sum_{j_\nu,m_\nu,n_\nu,\iota}\varphi^{j_\nu,\iota}_{m_\nu}(\phi)\;\mathcal{I}^{j_\nu,\iota}_{n_\nu}\prod_{i=1}^4 \sqrt{d_{j_{i}}}D^{j_{i}}_{m_{i},n_{i}}(g_{i}).
\ee
where $D^{j}_{m,n}(g)$ are the Wigner functions, $d_{j}$ is the dimension of the corresponding irreducible representation, \emph{i.e.} $d_j=2j+1$. The representation label $j$ takes integer and half-integer values, \emph{i.e.} $j\in\{0,\frac{1}{2},1,\frac{3}{2},\dots\}$, the indices $m$, $n$ take the values $-j\leq m,n\leq j$. Furthermore, the right-invariance leads more precisely to the Hilbert space $\mathcal{H}=L^2\left(\mbox{SU(2)}^4/\mbox{SU(2)},\de\mu_{Haar}\right)$. This is the \emph{intertwiner space} of a four-valent open spin network vertex, and also the Hilbert space of states for a single tetrahedron, a basis for which is given by the intertwiners $\mathcal{I}^{j_\nu,\iota}_{n_\nu}$, which are elements in 
\be\label{eq:HilbertSpace4Vertex}
\accentset{\circ}{\mathcal{H}}_{j_\nu}=\mbox{Inv}_{\mbox{\scriptsize SU(2)}}\left[\mathcal{H}_{j_1}\otimes\mathcal{H}_{j_2}\otimes\mathcal{H}_{j_3}\otimes\mathcal{H}_{j_4}\right].
\ee

The index $\iota$ labels elements in a basis in $\accentset{\circ}{\mathcal{H}}_{j_\nu}$, and represents an additional degree of freedom in the kinematical description of the GFT field and of its quantum states.

For example, $\iota$ can be chosen so as to label eigenstates of the volume operator for a single tetrahedron. With this choice, the volume operator acts diagonally on a wavefunction for a single tetrahedron $\varphi$ (which we indicate with the same symbol as the classical GFT field, since they are functionally analogous) decomposed as in Eq.~(\ref{eq:GFTfield})
\be\label{eq:VolumeActionGFT}
\hat{V}\varphi (g_\nu)=\sum_{j_\nu,m_\nu,n_\nu,\iota}V^{j_\nu,\iota}\varphi^{j_\nu,\iota}_{m_\nu}\;\mathcal{I}^{j_\nu,\iota}_{n_\nu}\prod_{i=1}^4 \sqrt{d_{j_{i}}}D^{j_{i}}_{m_{i},n_{i}}(g_{i}).
\ee
This action of the volume operator is the same as in LQG, where the volume eigenvalues for four-valent vertices have been studied extensively, see \emph{e.g.} Ref.~\cite{Brunnemann:aa}. 
More details on the quantum geometry of GFT states are in Appendix~\ref{sec:Volume}. 

The number of quanta (at a given clock time $\phi$) can be defined as a second quantised operator following Ref.~\cite{Oriti:2013aqa}. Its mean value in a coherent state of the field operator (the simplest type of condensate state) is the squared norm of the mean field $\varphi$
\be\label{eq:NumberOfQuanta}
N(\phi)=\int_{\rm SU(2)^4}\de\mu_{\textrm{Haar}}\;\overline{\varphi}(g_{\nu};\phi)\varphi(g_{\nu};\phi).
\ee
It is worth stressing that $N$ is in general not conserved by the dynamics in GFT. In fact, this is one of the main advantages of this approach, since it allows us to study efficiently the dynamics of quantum gravity states with a variable number of degrees of freedom.

\subsection{Group field theory for the Lorentzian EPRL model}\label{sec:2}
In the following, we work with the GFT formulation of the Lorentzian EPRL model for quantum gravity, developed first in the context of spin foam models. This was also used in Ref.~\cite{\BlockRefJ}, in a slightly generalised form, to account for some of the ambiguities in the definition of the model, including quantization ambiguities, other choices at the level of model building, and possible quantum corrections due to renormalization flow. It was presented using the spin representation of the GFT field $\varphi$ (thus of the action), and using the $\mbox{SU(2)}$ projection of the Lorentz structures in terms of which the model is originally defined. Like any GFT model, the action is decomposed into the sum of a kinetic and an interaction term
\be\label{eq:Action}
S=K+V_5+\overline{V}_5\, .
\ee
The most general (local) kinetic term for an $\mbox{SU(2)}$-based GFT field of rank-4 is\footnote{We indicate $\overline{\varphi^{j^{1}_\nu \iota_1}_{m^1_\nu}}$ by $\overline{\varphi}^{j^{1}_\nu \iota_1}_{m^1_\nu}$, for typographic reasons.}
\be\label{eq:kinetic}
K=\int\de\phi\;\sum_{j^a_{\nu},m^a_{\nu},\iota_a}\overline{\varphi}^{j^{1}_\nu\, \iota_1}_{m^1_\nu}\mathcal{K}^{j^1_\nu \,\iota_1}_{m^1_\nu}\varphi^{j^2_\nu\, \iota_2}_{m^2_\nu}\delta^{j^1_\nu j^2_\nu}\delta_{m^1_\nu m^2_\nu} \delta^{\iota_1 \iota_2},
\ee
and one has the interaction term corresponding to simplicial combinatorial structures given by\footnote{The authors would like to thank Marco Finocchiaro for pointing out an incorrect expression of the interaction potential that appeared in previous literature (private communication). The potential given here correctly implements the closure constraint of the 4-simplex obtained by `gluing' five tetrahedra.}
\be\label{eq:potential}
\begin{split}
V=\frac{1}{5}\int\de\phi\;\sum_{j_i,m_i,\iota_a}&\varphi^{j_1j_2j_3j_4\iota_1}_{m_1m_2m_3m_4}\varphi^{j_4j_5j_6j_7\iota_2}_{-m_4m_5m_6m_7}\varphi^{j_7j_3j_8j_9\iota_3}_{-m_7-m_3m_8m_9}\varphi^{j_9j_6j_2j_{10}\iota_4}_{-m_9-m_6-m_2m_{10}}\varphi^{j_{10}j_8j_5j_1\iota_5}_{-m_{10}-m_8-m_5-m_1}\\&\times\prod_{i=1}^{10}(-1)^{j_i-m_i}~\mathcal{V}_5(j_1,\dots,j_{10};\iota_1, \dots\iota_5) \; .
\end{split}
\ee
The details of the EPRL model would be encoded in the choice of kernels $\mathcal{K}$ and $\mathcal{V}_5$, and it is the interaction kernel that encodes the Lorentzian embedding of the theory and its full covariance, and what goes usually under the name of ``spin foam vertex amplitude'', here with boundary $\mbox{SU(2)}$-states. The explicit expression for such interaction kernel can be found in Ref.~\cite{\BlockRefJ} and, in more details in Ref.~\cite{Speziale:2016axj}. We do not need to be explicit about the functional form of the interaction kernel, in this paper, while we will say more about the kinetic term in the following.
Some discrete symmetries of the interaction kernel will however be relevant for what follows. In fact, the coefficients $\mathcal{V}_5$ are invariant under permutations of the spins and of the intertwiners,  which preserve the combinatorial structure of the potential (\ref{eq:potential}).

Beside the general form of Eq.~(\ref{eq:kinetic}), in the following we will also use a specific case for the GFT kinetic term, i.e. 
      \be\label{eq:freeaction}
   K=\int\de\phi\;\int_{\rm SU(2)^4}\de\mu_{\textrm{Haar}}\;\overline{\varphi}(g_{\nu},\phi)\mathcal{K}_{g_{\nu}}\varphi(g_{\nu},\phi)\; , 
   \ee
in the group representation, with
\be\label{eq:LocalKineticKernel}
\mathcal{K}_{g_{\nu}}=-\Big(\tau\pa_{\phi}^2+\sum_{i=1}^4\triangle_{g_i}\Big)+m^2 \; \;\; \tau, m^2 \, \in \mathbb{R},
\ee
which has been previously studied in the context of GFT cosmology in Ref.~\cite{Pithis:2016wzf,Pithis:2016cxg}, motivated by the renormalisation group analysis of GFT models (see Ref.~\cite{Carrozza:2013wda} and references therein). The same term can be given in the spin representation (using also the orthogonality of the intertwiners) as
\be\label{eq:freeactionSpin}
K=\int\de\phi\;\sum_{j_{\nu},m_{\nu},n_{\nu},\iota_1,\iota_2}\overline{\mathcal{I}}^{j_\nu \iota_1}_{n_{\nu}}\mathcal{I}^{j_\nu \iota_2}_{n_{\nu}}\overline{\varphi}^{j_{\nu}\iota_1}_{m_{\nu}}\hat{T}_{j_\nu}\varphi^{j_{\nu}\iota_2}_{m_{\nu}}=\int\de\phi\;\sum_{j_{\nu},m_{\nu},\iota}\overline{\varphi}^{j_{\nu}\iota}_{m_{\nu}}\hat{T}_{j_\nu}\varphi^{j_{\nu}\iota}_{m_{\nu}},
\ee
with
\be\label{eq:Toperator}
\hat{T}_{j_\nu}=-\tau\pa_{\phi}^2+\eta\sum_{i=1}^4j_i(j_i+1)+m^2.
\ee
Let us stress once more that the exact functional dependance on the discrete geometric data can be left more general for the EPRL model(s), since it is not uniquely fixed in the construction of the model, and it is only weakly constrained (mainly at large volumes) by the effective cosmological dynamics (which of course allows for the specific example above); the dependance on the scalar field variable $\phi$ is more important for obtaining the correct cosmological dynamics, at least in the isotropic case. More precisely, the choice of Eq.~(\ref{eq:freeactionSpin}) is a special case of the generalised EPRL model used in Ref.~\cite{\BlockRefJ}, obtained by the identification of the general functions used to parametrize the ambiguities in the EPRL model as $A_j=-\tau$, $B_j=-\big(4\eta j(j+1)+m^2\big)$. The quantity $m_j^2$, introduced in those references, is defined as $m_j^2=\frac{B_j}{A_j}$. In this case it will be given by the ratio $\frac{3 \eta +m^2}{\tau}$. At late times, this has to approximately coincide with the Newton's constant, in order to reproduce the Friedmann equations, as we will discuss in the next section. Notice that this will be a -definition- of Newton's constant from the fundamental parameters of the theory.

\subsection{Emergent Friedmann dynamics}\label{sec:Friedmann}
In this Section we derive the equations of motion of an isotropic cosmological background from the dynamics of the mean field for the GFT model. We reproduce in more detail the analysis of Ref.~\cite{\BlockRefJ} and clearly spell out all the assumptions made in the derivation, including the necessary restrictions on the GFT field, such as isotropy (\emph{i.e.} considering equilateral tetrahedra) and left-invariance, which we now discuss. 
\subsubsection{Conditions on the mean field}\label{sec:LeftInv}
As we have recalled, the simplest effective cosmological dynamics is obtained as the mean field approximation of the full GFT quantum theory, for any specific model. The resulting hydrodynamic equations, which we will discuss in the next section, are basically the classical equations of motion of the given GFT model, subject to a few additional restrictions. One way to obtain such equations from the microscopic quantum dynamics is to consider operator equations of motion evaluated in mean value on simple field coherent states, i.e. simple condensate states, and the resulting equations are going to be non-linear equations for the condensate wavefunctions, playing the role of classical GFT field. Such condensates have a geometric interpretation as homogeneous continuum spatial geometries and the condensate wavefunction is a probability distribution (a fluid density) on the space of such homogeneous geometries (i.e. minisuperspace, or the corresponding conjugate space).

\

For this interpretation to be valid, however, one additional condition has to be imposed on the condensate wave function: {\it left-invariance} under the diagonal group action. If the wavefunction satisfies this additional condition, on top of the right-invariance, and the domain is chosen to be $\mbox{SU(2)}^4$, coming from the imposition of simplicity constraints on $SL(2,\mathbb{C})$ data, like in the EPRL model, then the domain becomes isomorphic to minisuperspace of homogeneous geometries \cite{Gielen:2014ila}. Thus, this is not a symmetry of the GFT field, like the right-invariance, nor it is normally imposed on GFT quantum states. It is a property imposed on this specific class of states, in order to reduce the number of dynamical degrees of freedom and to guarantee the above interpretation.
Left-invariance leads to the following decomposition of the field components:
\be\label{eq:LeftInvarianceAnsatz}
\varphi^{j_{\nu}\iota}_{m_{\nu}}=\sum_{\iota^{\prime}}\varphi^{j_{\nu}\iota\iota^{\prime}}\mathcal{I}^{j_{\nu}\iota^{\prime}}_{m_{\nu}},
\ee
where $\iota^{\prime}$ is another intertwiner label, independent from $\iota$.  
Then, Eq.~(\ref{eq:kinetic}) becomes, using the assumption Eq.~(\ref{eq:LeftInvarianceAnsatz})
\be
K=\int\de\phi\;\sum_{j_{\nu},m_{\nu},\iota}\overline{\varphi}^{j_{\nu}\iota}_{m_{\nu}}\mathcal{K}^{j_\nu \iota}_{m_\nu}\varphi^{j_{\nu}\iota}_{m_{\nu}}=\int\de\phi\;\sum_{j_{\nu},\iota,\iota^{\prime},\iota^{\prime\prime}}\overline{\varphi}^{j_{\nu}\iota\iota^{\prime}}\tilde{\mathcal{K}}^{j_\nu \iota\iota^{\prime}\iota^{\prime\prime}}\varphi^{j_{\nu}\iota\iota^{\prime\prime}},
\ee
with
\be\label{eq:SimplificationGeneralLeftInv}
\tilde{\mathcal{K}}^{j_\nu \iota\iota^{\prime}\iota^{\prime\prime}}=\sum_{m_\nu}\overline{\mathcal{I}}^{j_{\nu}\iota^{\prime}}_{m_{\nu}}\mathcal{I}^{j_{\nu}\iota^{\prime\prime}}_{m_{\nu}}\mathcal{K}^{j_\nu \iota}_{m_\nu}\;\; , 
\ee
while the special case Eq.~(\ref{eq:freeactionSpin}) further simplifies to
\be\label{eq:freeactionSpinSimplified}
K=\int\de\phi\;\sum_{j_{\nu},m_{\nu},\iota}\overline{\varphi}^{j_{\nu}\iota}_{m_{\nu}}\hat{T}_{j_\nu}\varphi^{j_{\nu}\iota}_{m_{\nu}}=\int\de\phi\;\sum_{j_{\nu},\iota}\overline{\varphi}^{j_{\nu}\iota}\hat{T}_{j_\nu}\varphi^{j_{\nu}\iota}.
\ee
When the kernel $\mathcal{K}^{j_\nu \iota}_{m_\nu}$ does not depend on $m_\nu$, Eq.~(\ref{eq:SimplificationGeneralLeftInv}) simplifies considerably
\be
\tilde{\mathcal{K}}^{j_\nu \iota\iota^{\prime}\iota^{\prime\prime}}=\left(\sum_{m_\nu}\overline{\mathcal{I}}^{j_{\nu}\iota^{\prime}}_{m_{\nu}}\mathcal{I}^{j_{\nu}\iota^{\prime\prime}}_{m_{\nu}}\right)\mathcal{K}^{j_\nu \iota}=\delta^{\iota^{\prime}\iota^{\prime\prime}}\mathcal{K}^{j_\nu \iota}~,
\ee
leading us to the following expression for the kinetic term
\be
K=\int\de\phi\;\sum_{j_{\nu},\iota,\iota^{\prime}}\overline{\varphi}^{j_{\nu}\iota\iota^{\prime}}\mathcal{K}^{j_\nu \iota}\varphi^{j_{\nu}\iota\iota^{\prime}}~.
\ee
\

\subsubsection{Isotropic reduction and monochromatic tetrahedra}\label{sec:MaxVol}
As we discussed in the introduction, different definitions of \lq isotropy\rq~are possible for GFT condensates, and have been used in the literature, depending on the chosen reconstruction procedure for the continuum geometry out of the discrete data associated to such GFT states. For all of them, though, the result is qualitatively similar, as it should be: the condensate wavefunction has to depend on a single degree of freedom, e.g. one single spin variable, corresponding to the volume information or the scale factor of the emergent universe. Also, we do not expect that these different definitions of isotropic wavefunction would result in very different cosmological dynamics, for any given GFT model, and in fact this seems to be confirmed so far in the literature. In this paper, as in  Ref.~\cite{\BlockRefJ}, we adopt the simplest and most symmetric definition: we choose a condensate wavefunction such that the corresponding GFT quanta can be interpreted as equilateral tetrahedra. This implies that all of the spins labelling the quanta are equal $j_i=j, \;\forall i$, corresponding to tetrahedra with all triangle areas being equal. In this case, spin network vertices are said to be monochromatic.
We further assume that the only non-vanishing coefficients for each $j$ are those which correspond to the largest eigenvalue of the volume and a fixed orientation of the vertex (which lifts the degeneracy of the volume eigenvalues). In this way, the label $\iota$ is uniquely determined in each intertwiner space following from right-invariance (see Appendix~\ref{sec:Volume}). We call this particular value $\iota^{\star}$. This means that we have fixed all the quantum numbers of a quantum tetrahedron. 
We are still left with the intertwiner label $\iota^{\prime}$ following from left-invariance. To fix this, we identify the two vectors in $\accentset{\circ}{\mathcal{H}}_{j_\nu}$ determined by the decomposition of $\varphi$ by assuming that the only non-vanishing components $\varphi^{j_{\nu}\iota\iota^{\prime}}$ are such that $\iota=\iota^{\prime}$, \emph{i.e.}
\be\label{eq:LeftRightIdentify}
\varphi^{j_{\nu}\iota\iota^{\prime}}=\varphi^{j_{\nu}\iota\iota} \delta^{\iota\iota^{\prime}}~\hspace{1em} \mbox{(no sum)}.
\ee
Thus, also this extra label is fixed by the maximal volume requirement. The geometric interpretation of this further step is unclear, at present, but it is at least compatible with what we know about the (quantum) geometry of GFT states.
Using Eqs.~(\ref{eq:LeftInvarianceAnsatz}),~(\ref{eq:LeftRightIdentify}), the expansion Eq.~(\ref{eq:GFTfield}) simplifies to
\be\label{eq:LeftRightInvField}
\varphi(g_\nu;\phi)=\sum_{j_\nu,m_\nu,n_\nu,\iota}\varphi^{j_{\nu}\iota\iota}(\phi)\;\mathcal{I}^{j_{\nu}\,\iota}_{m_{\nu}}\mathcal{I}^{j_\nu\,\iota}_{n_\nu}\prod_{i=1}^4 \sqrt{d_{j_{i}}}D^{j_{i}}_{m_{i},n_{i}}(g_{i}).
\ee

Bringing all these conditions together (and dropping unnecessary repeated intertwiner labels), we get for the kinetic term
\be\label{eq:KineticSimplifiedABC}
K=\int\de\phi\;\sum_{j}\overline{\varphi}^{j\iota^{\star}}\tilde{\mathcal{K}}^{j\,\iota^{\star}}\varphi^{j\iota^{\star}}
\; , 
\ee
while the interaction term is given by
\be\label{eq:VwithVdoubleprime}
V=\frac{1}{5}\int\de\phi\;\sum_{j}\left(\varphi^{j\iota^\star}\right)^5\mathcal{V}^{\,\prime\prime}_5(j;\iota^\star).
\ee
In Eq.~(\ref{eq:VwithVdoubleprime}) we introduced the notation
\be\label{eq:IsotropicPotential}
\mathcal{V}^{\,\prime\prime}_5(j;\iota^\star)=\mathcal{V}^{\,\prime}_5(\underbrace{j,j\dots,j}_{10};\underbrace{\iota^\star,\iota^\star \dots\iota^\star}_{5})=
\mathcal{V}_5(j\dots j;\iota^\star \dots\iota^\star)\omega(j,\iota^\star),
\ee
with
\be\label{eq:OmegaDefinition}
\omega(j,\iota^\star)=\sum_{m_i}\prod_{i=1}^{10}(-1)^{j_i-m_i}~\mathcal{I}^{j\iota^\star}_{m_1m_2m_3m_4}\mathcal{I}^{j\iota^\star}_{-m_4m_5m_6m_7}\mathcal{I}^{j\iota^\star}_{-m_7-m_3m_8m_9}\mathcal{I}^{j\iota^\star}_{-m_9-m_6-m_2m_{10}}\mathcal{I}^{j\iota^\star}_{-m_{10}-m_8-m_5-m_1}.
\ee
Thus, in the isotropic case, the effect of the interactions is contained in the diagonal of the potential and in the coefficient $\omega(j,\iota^\star)$ constructed out of the intertwiners. We also observe that distinct monochromatic components have independent dynamics.

Eq.~(\ref{eq:OmegaDefinition}) can also be written in a different form, by expressing the intertwiner $\mathcal{I}^{j\iota^\star}_{m_\nu}$ in terms of the intertwiners $\alpha^{jJ}_{m_\nu}$ defined in Appendix~\ref{sec:Intertwiner} (see Eq.~(\ref{eq:DefAlphaIntertwiner}))
\be
\omega(j,\iota^\star)=\sum_{J_k}\left(\prod_{k=1}^5c^{J_k\iota^{\star}}\right)\{15j\}_{J_k}~,
\ee
where we identified the contraction of five intertwiners $\alpha^{jJ}_{m_\nu}$ with a 15j symbol of the first type
\be
\{15j\}_{J_k}=\sum_{m_i}\prod_{i=1}^{10}(-1)^{j_i-m_i}~\alpha^{jJ_1}_{m_1m_2m_3m_4}\alpha^{jJ_2}_{-m_4m_5m_6m_7}\alpha^{jJ_3}_{-m_7-m_3m_8m_9}\alpha^{jJ_4}_{-m_9-m_6-m_2m_{10}}\alpha^{jJ_5}_{-m_{10}-m_8-m_5-m_1}~.
\ee

\subsubsection{Background equation}\label{sec:Background}
The equations of motion for the background can be found by varying the action Eq.~(\ref{eq:Action}). Using Eqs.~(\ref{eq:KineticSimplifiedABC}),~(\ref{eq:IsotropicPotential}) we find (compare with Ref.~\cite{\BlockRefJ})
\be\label{eq:Background}
\tilde{\mathcal{K}}^{j\,\iota^{\star}}\varphi^{j\iota^{\star}}+\mathcal{V}^{\,\prime\prime}_5(j;\iota^\star) \left(\varphi^{j\iota^\star}\right)^4=0.
\ee
In the particular case given by Eq.~(\ref{eq:LocalKineticKernel}) we can write
\be
K=\int\de\phi\;\sum_{j}\overline{\varphi}^{j\iota^{\star}}\hat{T}_{j}\varphi^{j\iota^{\star}} \qquad 
\hat{T}_{j}=-\tau\pa_{\phi}^2+4\eta j(j+1)+m^2.
\ee
For the purpose of studying a concrete example, from now on we consider the special case in which $j=\frac{1}{2}$ of $\mbox{SU(2)}$. Then, we have, using the definition Eq.~(\ref{eq:OmegaDefinition})
\be\label{eq:OmegaVolumeEigenstatePlus}
\omega\left(\frac{1}{2},\pm\right)=\sum_{m_i}\mathcal{I}^{\frac{1}{2}\,\iota_\pm}_{m_1m_2m_3m_4}\mathcal{I}^{\frac{1}{2}\,\iota_\pm}_{m_4m_5m_6m_7}\mathcal{I}^{\frac{1}{2}\,\iota_\pm}_{m_7m_3m_8m_9}\mathcal{I}^{\frac{1}{2}\,\iota_\pm}_{m_9m_6m_2m_{10}}\mathcal{I}^{\frac{1}{2}\,\iota_\pm}_{m_{10}m_8m_5m_1}=\frac{3\mp i \sqrt{3}}{18 \sqrt{2}}.
\ee
$\iota^\star=\iota_\pm$ means that we are considering as an intertwiner the volume eigenvector corresponding to a positive (resp. negative) orientation, see Appendix~\ref{sec:Volume}.

Thus, the equation of motion for the background in this special case reads
\be\label{eq:BackgroundDynamics}
\left(-\tau\pa_{\phi}^2+3\eta+m^2\right)\varphi^{\frac{1}{2}\iota^\star}+\overline{\mathcal{V}^{\,\prime\prime}}_5\left(\frac{1}{2};\iota^\star\right)\left(\overline{\varphi}^{\frac{1}{2}\iota^\star}\right)^4=0,
\ee
with the coefficient of the interaction term given by Eqs.~(\ref{eq:IsotropicPotential}),~(\ref{eq:OmegaVolumeEigenstatePlus}). Notice that, under the assumption that $j=\frac{1}{2}$, used in addition to the isotropic reduction, even the more general form of the EPRL GFT model coupled to a massless free scalar field, Eq.~(\ref{eq:kinetic}), as used in  Ref.~\cite{\BlockRefJ} will collapse to the special case Eq.~(\ref{eq:BackgroundDynamics}), for some constant $\eta$. Also, the dominance of a single (small) spin component in the cosmological dynamics of isotropic backgrounds can be shown to take place at late times Ref.~\cite{Gielen:2016uft}, and it can be expected to be a decent approximation at earlier ones. Thus, even the special case we are considering is not too restrictive.
For the many results obtained about the above effective cosmological dynamics, which can be turned into an evolution equation (in relational time) for the volume of the universe, we refer to the literature. 
These include: an emergent Friedmann equation at late times \cite{\BlockRefJ}, a generic quantum bounce replacing the big bang singularity and a subsequent acceleration \cite{deCesare:2016axk}, several studies on the effects of GFT interactions \cite{Pithis:2016wzf,Pithis:2016cxg} (which become dominant away from the bounce), including a prolonged phase of acceleration and a later re-collapse of the universe to produce a cyclic evolution \cite{deCesare:2016rsf}, the relation with LQC dynamics \cite{Calcagni:2014tga,\BlockRefJ}.

Next, we focus our attention on perturbations around the isotropic case described in this section and governed by the above equations. 

\section{Non-monochromatic perturbations}\label{sec:Perturbations}
We can now derive the equations of motion for perturbations around an isotropic background GFT field configuration $\varphi_0$ satisfying Eq.~(\ref{eq:BackgroundDynamics}). 
\be
\varphi=\varphi_0+\delta\varphi.
\ee
Let us start by writing down the more general equation of motion for the background, by relaxing the isotropy assumption while retaining the other hypotheses of Sections~\ref{sec:LeftInv}.
\be
0=\frac{\delta S[\varphi]}{\delta\overline{\varphi}^{abcd\,\tilde{\iota}}}=\frac{\delta K[\varphi]}{\delta\overline{\varphi}^{abcd\,\tilde{\iota}}}+\frac{\delta \overline{V}_5[\varphi]}{\delta\overline{\varphi}^{abcd\,\tilde{\iota}}}
\ee
Above, we wrote explicitly all of the spin labels $j_\nu=(a,b,c,d)$. The first term is equal to
\be
\frac{\delta K[\varphi]}{\delta\overline{\varphi}^{abcd\,\tilde{\iota}}}=\mathcal{K}^{abcd\,\tilde{\iota}}\varphi^{abcd\,\tilde{\iota}} \; \; ,
\ee
while for the second term we have
\be\label{eq:InteractionTermGeneralEOM}
\begin{split}
\frac{\delta V_5[\varphi]}{\delta\varphi^{abcd\,\tilde{\iota}}}&=\frac{1}{5}\sum \left[ \varphi^{d567\,\iota_2}\varphi^{7c89\,\iota_3}\varphi^{96b10\,\iota_4}\varphi^{1085a\,\iota_5}\mathcal{V}^{\,\prime}_5\left(\scriptstyle{a,b,c,d,5,6,7,8,9,10;\tilde{\iota},\iota_2,\iota_3,\iota_4,\iota_5}\right)+\right.\\
&\phantom{=\frac{1}{5}\sum [ } \varphi^{123a\,\iota_1}\varphi^{d389\,\iota_3}\varphi^{9c210\,\iota_4}\varphi^{108b1\,\iota_5}\mathcal{V}^{\,\prime}_5\left(\scriptstyle{1,2,3,a,b,c,d,8,9,10;\iota_1,\tilde{\iota},\iota_3,\iota_4,\iota_5}\right)+\\
&\phantom{=\frac{1}{5}\sum [ }\varphi^{12b4\,\iota_1}\varphi^{456a\,\iota_2}\varphi^{d6210\,\iota_4}\varphi^{10c51\,\iota_5}\mathcal{V}^{\,\prime}_5\left(\scriptstyle{1,2,b,4,5,6,a,c,d,10;\iota_1,\iota_2,\tilde{\iota},\iota_4,\iota_5}\right)+\\
&\phantom{=\frac{1}{5}\sum [ }\varphi^{1c34\,\iota_1}\varphi^{45b7\,\iota_2}\varphi^{738a\,\iota_3}\varphi^{d851\,\iota_5}\mathcal{V}^{\,\prime}_5\left(\scriptstyle{1,c,3,4,5,b,7,8,a,d;\iota_1,\iota_2,\iota_3,\tilde{\iota},\iota_5}\right)+\\
&\phantom{=\frac{1}{5}\sum [ }\left. \varphi^{d234\,\iota_1}\varphi^{4c67\,\iota_2}\varphi^{73b9\,\iota_3}\varphi^{962a\,\iota_4}\mathcal{V}^{\,\prime}_5\left(\scriptstyle{d,2,3,4,c,6,7,b,9,a;\iota_1,\iota_2,\iota_3,\iota_4,\tilde{\iota}}\right)\right].
\end{split}
\ee
By just relabelling the indices, we obtain
\be
\begin{split}
\frac{\delta V_5[\varphi]}{\delta\varphi^{abcd\,\tilde{\iota}}}=&
\frac{1}{5}\sum\varphi^{d567\,\iota_2}\varphi^{7c89\,\iota_3}\varphi^{96b10\,\iota_4}\varphi^{1085a\,\iota_5}[\mathcal{V}^{\,\prime}_5\left(\scriptstyle{a,b,c,d,5,6,7,8,9,10;\tilde{\iota},\iota_2,\iota_3,\iota_4,\iota_5}\right)+\\&\mathcal{V}^{\,\prime}_5\left(\scriptstyle{10,8,5,a,b,c,d,6,7,9;\iota_5,\tilde{\iota},\iota_2,\iota_3,\iota_4}\right)+\mathcal{V}^{\,\prime}_5\left(\scriptstyle{9,6,b,10,8,5,a,c,d,7;\iota_4,\iota_5,\tilde{\iota},\iota_2,\iota_3}\right)+\\&\mathcal{V}^{\,\prime}_5\left(\scriptstyle{7,c,8,9,6,b,10,5,a,d;\iota_3,\iota_4,\iota_5,\tilde{\iota},\iota_2}\right)+\mathcal{V}^{\,\prime}_5\left(\scriptstyle{d,5,6,7,c,8,9,b,10,a;\iota_2,\iota_3,\iota_4,\iota_5,\tilde{\iota}}\right)] \;\; ,
\end{split}
\ee
which becomes, taking into account the discrete symmetries of the interaction kernel, the simpler expression 
\be\label{eq:InteractionTermGeneral}
\frac{\delta V_5[\varphi]}{\delta\varphi^{abcd\,\tilde{\iota}}}=\sum\varphi^{d567\,\iota_2}\varphi^{7c89\,\iota_3}\varphi^{96b10\,\iota_4}\varphi^{1085a\,\iota_5}\mathcal{V}^{\,\prime}_5\left(\scriptstyle{a,b,c,d,5,6,7,8,9,10;\tilde{\iota},\iota_2,\iota_3,\iota_4,\iota_5}\right).
\ee

Moreover, given the structure of the interaction term in Eq.~(\ref{eq:InteractionTermGeneral}), the only non-vanishing contributions to the first order dynamics of the perturbations around a monochromatic background come from terms having at least three identical spins among $(a,b,c,d)$. Therefore, depending on which of the four indices, labeled $j^\prime$ is singled out to be different from the other three, labeled $j$, we obtain four independent equations
\be
\mathcal{K}^{jjjj^{\prime}\,\tilde{\iota}}\delta\varphi^{jjjj^{\prime}\,\tilde{\iota}}+\sum_{\iota}\delta\overline{\varphi}^{j^{\prime}jjj\,\iota}\left(\overline{\varphi_0}^{j\,\iota^\star}\right)^3\mathcal{V}^{\,\prime}_5\left(\scriptstyle{j,j,j,j^{\prime},j,j,j,j,j,j;\tilde{\iota},\iota,\iota^\star,\iota^\star,\iota^\star}\right)=0.
\ee
\be
\mathcal{K}^{jj j^{\prime}j\,\tilde{\iota}}\delta\varphi^{jj j^{\prime}j\,\tilde{\iota}}+\sum_{\iota}\delta\overline{\varphi}^{jj^{\prime}jj\,\iota}\left(\overline{\varphi_0}^{j\,\iota^\star}\right)^3\mathcal{V}^{\,\prime}_5\left(\scriptstyle{j,j,j^{\prime},j,j,j,j,j,j,j;\tilde{\iota},\iota^\star,\iota,\iota^\star,\iota^\star}\right)=0.
\ee
\be
\mathcal{K}^{j j^{\prime}jj\,\tilde{\iota}}\delta\varphi^{j j^{\prime}jj\,\tilde{\iota}}+\sum_{\iota}\delta\overline{\varphi}^{jjj^{\prime}j\,\iota}\left(\overline{\varphi_0}^{j\,\iota^\star}\right)^3\mathcal{V}^{\,\prime}_5\left(\scriptstyle{j, j^{\prime},j,j,j,j,j,j,j,j;\tilde{\iota},\iota^\star,\iota^\star,\iota,\iota^\star}\right)=0.
\ee
\be
\mathcal{K}^{j^{\prime}jjj\,\tilde{\iota}}\delta\varphi^{j^{\prime}jjj\,\tilde{\iota}}+\sum_{\iota}\delta\overline{\varphi}^{jjjj^{\prime}\,\iota}\left(\overline{\varphi_0}^{j\,\iota^\star}\right)^3\mathcal{V}^{\,\prime}_5\left(\scriptstyle{j^{\prime},j,j,j,j,j,j,j,j,j;\tilde{\iota},\iota^\star,\iota^\star,\iota^\star,\iota}\right)=0.
\ee
We define a new function
\be
\mathcal{U}(j,j^{\prime},\iota,\iota^{\prime};n)\equiv\left(\overline{\varphi_0}^{j\,\iota^\star}\right)^3\mathcal{V}^{\,\prime}_5\left(\scriptstyle{\underbrace{\scriptstyle{j,\dots,j^{\prime}}}_\text{n},\dots,j,j,j,j,j,j,j;\underbrace{\scriptstyle{\iota,\dots,\iota^{\prime}}}_{5-n},\dots,\iota^\star}\right),
\ee
with $j^{\prime}$ in the $n$-th position ($n=1,2,3,4$) and $\iota^{\prime}$ appearing in position $5-n$ after $\iota$, which keeps the first place. For instance, one has for $n=1$
\be\label{eq:DefinitionEffectivePotential}
\mathcal{U}(j,j^{\prime},\iota^\star,\iota,\iota^{\prime};n)=\left(\overline{\varphi_0}^{j\,\iota^\star}\right)^3\mathcal{V}^{\,\prime}_5\left(\scriptstyle{j^{\prime},j,j,j,j,j,j,j,j,j;\iota,\iota^\star,\iota^\star,\iota^\star,\iota^{\prime}}\right).
\ee
Thus, the equations of motion for the perturbations can be rewritten more compactly as
\begin{align}\label{eq:PerturbationsFirstIndex}
\mathcal{K}^{j^{\prime}jjj\,\iota}\delta\varphi^{j^{\prime}jjj\,\iota}+\sum_{\iota^{\prime}}\delta\overline{\varphi}^{jjjj^{\prime}\,\iota^{\prime}}\mathcal{U}(j,j^{\prime},\iota^\star,\iota,\iota^{\prime};1)&=0\nonumber\\
\mathcal{K}^{jj^{\prime}jj\,\iota}\delta\varphi^{jj^{\prime}jj\,\iota}+\sum_{\iota^{\prime}}\delta\overline{\varphi}^{jjj^{\prime}j\,\iota^{\prime}}\mathcal{U}(j,j^{\prime},\iota^\star,\iota,\iota^{\prime};2)&=0\nonumber\\
\mathcal{K}^{jjj^{\prime}j\,\iota}\delta\varphi^{jjj^{\prime}j\,\iota}+\sum_{\iota^{\prime}}\delta\overline{\varphi}^{jj^{\prime}jj\,\iota^{\prime}}\mathcal{U}(j,j^{\prime},\iota^\star,\iota,\iota^{\prime};3)&=0\nonumber\\
\mathcal{K}^{jjjj^{\prime}\,\iota}\delta\varphi^{jjjj^{\prime}\,\iota}+\sum_{\iota^{\prime}}\delta\overline{\varphi}^{j^{\prime}jjj\,\iota^{\prime}}\mathcal{U}(j,j^{\prime},\iota^\star,\iota,\iota^{\prime};4)&=0 \;\; .
\end{align}

With the particular kinetic kernel~(\ref{eq:LocalKineticKernel}), one has that the kinetic operator acting on the perturbation does not depend on the position of the perturbed index $j^{\prime}$, neither it depends on the intertwiner label $\iota$. Hence, in that case we can define
\be
\mathcal{K}^{\prime}=\mathcal{K}^{j^{\prime}jjj\,\iota}=\mathcal{K}^{jj^{\prime}jj\,\iota}=\mathcal{K}^{jjj^{\prime}j\,\iota}=\mathcal{K}^{jjjj^{\prime}\,\iota}=-\tau\pa_{\phi}^2+\eta\left(3j(j+1)+j^{\prime}(j^{\prime}+1)\right)+m^2.
\ee

The above equations are generic. However, recoupling theory imposes several restrictions on our perturbations, due to the conditions imposed on the fields: 
a)  $j^{\prime}$ is an integer (half-integer) if the background spin $j$ is an integer (half-integer); b) $j^{\prime}$ cannot be arbitrarily large, since for $j^{\prime}>3j$ the closure (right-invariance) condition would be violated; c) of course,  the case $j^{\prime}=j$ is uninteresting since such perturbations can be reabsorbed into the monochromatic background. 

In the simplest example $j=\frac{1}{2}$ there is only one permitted value for the perturbed spin, namely $j^{\prime}=\frac{3}{2}$, and the perturbation is identified with the state such that the total spin of a pair is $J=1$. Any such state is trivially also a volume eigenstate since the volume operator is identically vanishing in such intertwiner space, as it is one-dimensional (see Appendix~\ref{sec:Volume}, in particular the comment after Eq.~(\ref{eq:VolumeSpectrum})~).  For this reason, we will omit the indices $\iota$, $\iota^{\prime}$ in the following.

Let us introduce some further notation for these specific perturbations. We define
\be\label{eq:PerturbedIndices}
\psi_1=\delta\varphi^{\frac{3}{2}\frac{1}{2}\frac{1}{2}\frac{1}{2}}, \hspace{1em} \psi_2=\delta\varphi^{\frac{1}{2}\frac{3}{2}\frac{1}{2}\frac{1}{2}}, \hspace{1em} \psi_3=\delta\varphi^{\frac{1}{2}\frac{1}{2}\frac{3}{2}\frac{1}{2}}, \hspace{1em} \psi_4=\delta\varphi^{\frac{1}{2}\frac{1}{2}\frac{1}{2}\frac{3}{2}}
\ee
and similarly
\be
\mathcal{K}_1=\mathcal{K}^{\frac{3}{2}\frac{1}{2}\frac{1}{2}\frac{1}{2}}, \hspace{1em} \mathcal{K}_2=\mathcal{K}^{\frac{1}{2}\frac{3}{2}\frac{1}{2}\frac{1}{2}}, \hspace{1em}\mathcal{K}_3=\mathcal{K}^{\frac{1}{2}\frac{1}{2}\frac{3}{2}\frac{1}{2}}, \hspace{1em} \mathcal{K}_4=\mathcal{K}^{\frac{1}{2}\frac{1}{2}\frac{1}{2}\frac{3}{2}}.
\ee
Hence, it follows from Eq.~(\ref{eq:PerturbationsFirstIndex}) that the dynamics of the perturbations is governed (to first order) by the following equations (omitting the perturbation variables $j^{\prime}$, $\iota$, $\iota^{\prime}$ and the background spin $j=\frac{1}{2}$ in the argument of $\mathcal{U}$, Eq.~(\ref{eq:DefinitionEffectivePotential})):
\begin{align}
\mathcal{K}_1 \psi_1+\mathcal{U}(\iota^{\star};1)\overline{\psi_4}&=0 \nonumber \\
\mathcal{K}_4 \psi_4+\mathcal{U}(\iota^{\star};4)\overline{\psi_1}&=0\nonumber \\
\mathcal{K}_2 \psi_2+\mathcal{U}(\iota^{\star};2)\overline{\psi_3}&=0 \nonumber \\
\mathcal{K}_3 \psi_3+\mathcal{U}(\iota^{\star};3)\overline{\psi_2}&=0 \;\; \label{eq:System1}.
\end{align}
 %
 %
The resulting equations for the perturbations are reasonably simple, thanks mainly to the isotropy assumption on the background, which simplifies considerably the contribution from the GFT interaction term $U$. However, even the simplified functional form in which the Lorentzian EPRL vertex amplitude appears in these equations remains unknown in exact analytic terms. The above equations would have then to be studied numerically or in more phenomenological approach, in which the exact function $U$ is replaced by some simpler trial function, or several ones in different ranges of the variable $j$, approximating it. 
Luckily, for our present concerns, which relate to the behaviour of perturbations close to the cosmological bounce, these difficulties can be sidestepped since the interaction term is generically subdominant in that regime of the theory, mainly due to the smallness of the background condensate wavefunction (in turn related to the smallness of the universe 3-volume). This will allow us to perform a study of this dynamics in the following section.

Before we turn to such dynamics, let us notice that the equations (\ref{eq:System1}) make manifest an asymmetry of the interaction terms of the GFT model we are considering, more specifically of the EPRL vertex amplitude, that is not apparent at first sight. The equations in fact couple perturbations in the first field argument with perturbations in the fourth, and perturbations in the second with perturbations in the third, with no other combination being present. This happens despite the isotropy assumption on the background  and the other symmetries of the model. One can trace this asymmetry back to the combinatorial structure of the vertex amplitude itself: it corresponds to a 4-simplex as projected down to the plane but it is not symmetric with respect to the face pairings, if such faces are ordered in their planar projection: it only couples first and fourth faces across common tetrahedra sharing them, or second and third ones, i.e. exactly the type of asymmetry that is revealed in our perturbations equations. It is tempting to relate this asymmetry to an issue with orientability of the triangulations resulting from the Feynman expansion of the model, since the same type of issue has been identified in the Boulatov model for 3d gravity in Ref.~\cite{Freidel:2009hd}. It is unclear at this stage whether this is a problem or just a feature of the model; it is also unclear, in case one decides to remove such asymmetry, what is the best way to do so. The strategy followed in the 3d case  Ref.~\cite{Freidel:2009hd}, \emph{i.e.} to maintain the ordering of the GFT field arguments but modify the combinatorics of the interaction vertex to ensure orientability, does not seem available in this 4d case. An easy solution would be to impose that the GFT fields themselves are invariant under (even) permutations of their arguments, which also ensure orientability of the resulting triangulations. We leave this point, not directly relevant for the analysis of the next section, for further study.

\section{Dynamics of the perturbations at the bounce}\label{sec:DynamicsAtBounce}
We now study the dynamics of the perturbations around a background homogeneous and isotropic solution of the fundamental QG dynamics, in the mean field approximation.
We focus on the bounce regime, since this is where typical bouncing models of the early universe have difficulties in controlling the dynamics of anisotropies. 
Luckily, as anticipated, this is also the regime where, in the GFT condensate cosmology framework we can have the best analytic control over the (quantum) dynamics of the theory, at least in the mean field approximation. 
In fact, the bouncing regime takes place, in the hydrodynamic approximation we are working in, for low densities, thus, intuitively, for low values of the modulus of the GFT mean field. 

Considering the kernel of Eq.~(\ref{eq:LocalKineticKernel}), assuming $j=\frac{1}{2}$ for the background, and neglecting the interaction term, the background equation reads as
\be\label{eq:BackgroundBounceFundRepr}
\left(-\tau\pa_{\phi}^2+3\eta+m^2\right)\varphi^{\frac{1}{2}\iota^\star}\simeq0.
\ee
On the other hand, to first order, perturbations satisfy the equation 
\be\label{eq:EquationPerturbationNoInteractions}
\mathcal{K}^{\prime}\psi\simeq0,
\ee
where
\be
\mathcal{K}^{\prime}=-\tau\pa_{\phi}^2+\eta\left(3j(j+1)+j^{\prime}(j^{\prime}+1)\right)+m^2=-\tau\pa_{\phi}^2+6\eta+m^2 \; ,
\ee
and we have indicated a generic perturbation by $\psi$, since there is no difference among them, in this approximation.

When the interaction term is no longer subdominant (\emph{i.e.} after the Universe exits the bouncing phase and after it has expanded enough), the dynamics of the perturbations is given by the systems of equations (\ref{eq:System1}), which remain valid until $\psi\simeq\varphi^{\frac{1}{2}\iota^\star}$. At that point, higher order corrections are needed. On the other hand, it is important to stress that since the equations of motion become linear  at the bounce, at that point we are no longer subject to the constraint that non-monochromatic components should be small. In other words, $\psi\simeq\varphi^{\frac{1}{2}\iota^\star}$ is allowed in that regime and perturbations can be large. This observation will be important in the following.

Using the analytic expression of the background solution, given in Ref.~\cite{deCesare:2016axk}, we have
\be\label{eq:AnalyticBackground}
|\varphi^{\frac{1}{2}\iota^\star}|=\frac{e^{\sqrt{\frac{3 \eta +m^2}{\tau} } (\Phi-\phi )}
   \sqrt{-2 E_0 e^{2 \sqrt{\frac{3 \eta +m^2}{\tau} } (\phi
   -\Phi)} \sqrt{\Omega_0}+e^{4 \sqrt{\frac{3 \eta +m^2}{\tau} } (\phi
   -\Phi)} \Omega_0+\Omega_0}}{2 \sqrt{\frac{3 \eta +m^2}{\tau} } \sqrt[4]{\Omega_0}},
\ee
where
\be\label{eq:OmegaParameterBackground}
\Omega_0=E_0^2+4 Q_0^2 \left(\frac{3\eta +m^2}{\tau} \right)
\ee
and $E_0$, $Q_0$ are conserved quantities\footnote{The conservation of $Q_0$ is not exact, as it follows from an approximate $U(1)$-symmetry, which holds as long as interactions are negligible.}. $E_0$ is referred to as the ``GFT energy'' \cite{\BlockRefJ} of the monochromatic background\footnote{The \lq GFT energy\rq~is the total mechanical energy of an associated one-dimensional mechanical problem which governs the evolution of the modulus of the GFT mean field. Its relation to any macroscopic conserved quantity is not known at this stage.}.
Reality of the expression in Eqs.~(\ref{eq:AnalyticBackground}) implies
\be
\Omega_0\geq 0.
\ee
In order to have the same dynamics for the background as in Refs.~\cite{\BlockRefJ,deCesare:2016axk}, we demand that
\be\label{eq:BackgroundInequality}
\frac{3 \eta +m^2}{\tau}>0.
\ee
In this case, the modulus of the backround $|\varphi^{\frac{1}{2}\iota^\star}|$ has a unique global minimum at $\phi=\Phi$, corresponding to the quantum bounce.
We will consider two possible cases in which condition (\ref{eq:BackgroundInequality}) is still satisfied, but different conditions are imposed on the parameters governing the dynamics of the perturbations. This gives qualitatively the same evolution of the background but two radically different pictures for the evolution of the perturbations.

\

\begin{itemize}
\item Case \textit{i})~The first possibility is that $\tau,~ m^2 \geq0$. In this case, also the perturbations satisfy an analogous condition
\be\label{eq:PerturbationInequality}
\frac{6 \eta +m^2}{\tau}>0.
\ee
The analytic solution of the equation for the perturbations has the same form as Eq.~(\ref{eq:AnalyticBackground})
\be\label{eq:AnalyticPerturbations}
|\psi|=\frac{e^{\sqrt{\frac{6 \eta +m^2}{\tau} } (\Phi_1-\phi )}
   \sqrt{-2 E_1 e^{2 \sqrt{\frac{6 \eta +m^2}{\tau} } (\phi
   -\Phi_1)} \sqrt{\Omega_1}+e^{4 \sqrt{\frac{6 \eta +m^2}{\tau} } (\phi
   -\Phi_1)} \Omega_1+\Omega_1}}{2 \sqrt{\frac{6 \eta +m^2}{\tau} } \sqrt[4]{\Omega_1}}.
\ee
We introduced the quantity $\Omega_1$, in analogy with Eq.~(\ref{eq:OmegaParameterBackground})
\be\label{eq:DefOmega1}
\Omega_1=E_1^2+4 Q_1^2 \left(\frac{6\eta +m^2}{\tau} \right).
\ee
$E_1$ and $Q_1$ are two conserved quantities. We will refer to $E_1$ as to the ``GFT energy'' of the perturbations. Reality of Eq.~(\ref{eq:AnalyticPerturbations}) requires that $\Omega_1\geq0$. $|\psi|$ has a minimum at $\Phi_1$.
From Eqs.~(\ref{eq:AnalyticBackground}),~(\ref{eq:AnalyticPerturbations}), we find in the limit of large $\phi$
\be
\frac{|\psi|}{|\varphi^{\frac{1}{2}\iota^\star}|}\sim e^{ \Big(\sqrt{\frac{6 \eta +m^2}{\tau}} - \sqrt{\frac{3 \eta +m^2}{\tau}}\Big)\phi},
\ee
which means that perturbations cannot be neglected in this limit, \emph{i.e.} away from the bounce occurring at $\phi=\Phi$ (see Eq.~(\ref{eq:AnalyticBackground}) and discussion below Eq.~(\ref{eq:BackgroundInequality})). Therefore, when we are in this region of parameter space, they should be properly taken into account. Depending on the values of the parameters, they can become dominant already close to the bounce. At the same time, the value of the "time" $\phi$ at which this approximation is usable cannot be too large, because then we expect the GFT interactions to grow in importance, breaking the approximation on the background and perturbation dynamics we have assumed to be valid so far.

\item Case \textit{ii})~A second possibility is represented by the case in which condition (\ref{eq:BackgroundInequality}) is still satisfied while inequality (\ref{eq:PerturbationInequality}) is not. This can be accomplished with $\tau<0$ and $-6\eta<m^2<-3\eta$. In this case, the modulus of the perturbations oscillates around the minimum of a one-dimensional mechanical potential. 

Writing $|\psi|=\rho$, its dynamics is given by (see Refs.~\cite{\BlockRefJ,deCesare:2016axk,deCesare:2016rsf})
\be\label{eq:RadialPartPerturbations}
\partial_{\phi}^2 \rho=-U_{,\rho},
\ee
where
\be\label{eq:PotentialMechanics}
U(\rho)=\frac{Q^2_1}{2\rho^2}-\left(\frac{6\eta+m^2}{\tau}\right)\frac{\rho^2}{2}.
\ee
What happens in this case is that, away from the bounce, the perturbations are always dominated by the background.

In order to see this in a more quantitative way, we can make the simplifying assumption that the minimum of $U$ and the amplitude of the oscillations of $\rho$ are such that the interactions between quanta are always negligible for the perturbations. This can be realised by making an appropriate choice for the values of the parameters of the model. 

In this case, Eq.~(\ref{eq:RadialPartPerturbations}) describes the evolution of the perturbations at all times. Their qualitative behaviour around the bounce is illustrated in Fig.~\ref{Fig:AtoV}. Non-monochromatic perturbations are relevant at the bounce but drop off quickly away from it.\footnote{This is reminiscent of the results obtained in the context of a different model in Ref.~\cite{Pithis:2016cxg}, also suggesting that such non-monochromatic modes are only relevant in the regime of small volumes.} 

The behaviour of the perturbations is oscillatory, since $|\psi|$ is trapped in the potential well $U$ of Eq.~(\ref{eq:PotentialMechanics}). As a consequence of this, the number of non-monochromatic quanta $N_1=|\psi|^2$ has an upper bound. Conversely, the number of quanta in the background grows unbounded. 

We conclude that, in this window of parameter space, perturbations can be relevant at the bounce but are negligible for large numbers of quanta in the background. 
For a suitable strength of the interactions, non-monochromatic perturbations can become completely irrelevant for the dynamics before interactions kick in.

\end{itemize}

\subsection*{Measures of deviations from monochromatic GFT condensates}
It is interesting to explore further the deviations from perfect monochromaticity, by computing some quantities which can characterise the dynamics of the perturbed condensate and distinguish it from the purely monochromatic case. We do so in the following. The quantities we compute do not have a clear cosmological meaning, and do not correspond to specific gauge-invariant observables characterizing anisotropies in relativistic cosmology. They are however well-defined formal observables for GFT condensates. 

The first one we consider is a \emph{surface-area-to-volume ratio}. A first quantized area operator in GFT can be defined for a tetrahedron as in Ref.~\cite{Gielen:2015kua}: $\hat{A}=\kappa\sum_{i=1}^4\sqrt{-\Delta_i}$, where the sum runs over all the faces of the tetrahedron, in analogy with the LQG area operator. 

We have for its expectation value on a single monochromatic (equilateral) quantum:
\be
A_0=2\kappa\sqrt{3}
\ee
and for a perturbed non-monochromatic quantum:
\be
A_1=\frac{\kappa}{2}\sqrt{3}\left(3+\sqrt{5}\right) \qquad .
\ee

This operator can then be turned into a second quantised counterpart of the same (see \emph{e.g.} Ref.~\cite{\BlockRefE}), to be applied to ensembles of tetrahedra.  One can then easily compute the expectation value of this operator, as well as the expectation value of the total volume operator, in both an unperturbed and in a perturbed condensate state (of the simplest type considered in this paper). The resulting quantity is heuristically the sum of the areas of the four faces of each tetrahedron times the number of tetrahedra with the same areas.

The area-to-volume ratio for the example considered can then be expressed as
\be\label{eq:AreaToVolume}
\frac{A}{V}=\frac{A_0 N_0+A_1 N_1}{V_0 N_0}=\frac{A_0}{V_0}\left(1+\frac{A_1}{A_0}~\frac{N_1}{N_0}\right).
\ee
$A_0$ is the surface area of an unperturbed quantum and $A_1$ that of a perturbed one. $N_0$ and $N_1$ are the corresponding number of quanta, which can be computed using Eq.~(\ref{eq:NumberOfQuanta})
\begin{align}
N_0&=|\varphi^{\frac{1}{2}\iota^\star}|^2,\\
N_1&=|\psi|^2.
\end{align}
$V_0$ is the volume of a quantum of space in the background. We recall that the perturbed quanta considered in this example have vanishing volume (see Appendix~\ref{sec:Volume}). This has significant consequences which we illustrate in the following.
Hence, Eq.~(\ref{eq:AreaToVolume}) leads to
\be
\frac{A}{V}=\frac{A_0}{V_0}\left(1+\frac{3+\sqrt{5}}{4}~\frac{N_1}{N_0}\right).
\ee
Since $\frac{N_1}{N_0}\geq0$, we have
\be
\frac{A}{V}\geq\frac{A_0}{V_0}.
\ee
This inequality means that, for a given volume, quanta have \emph{on average} more surface than they would in a purely mono-chromatic (isotropic) background. 

\begin{figure}[b]
\includegraphics[width=0.6\columnwidth]{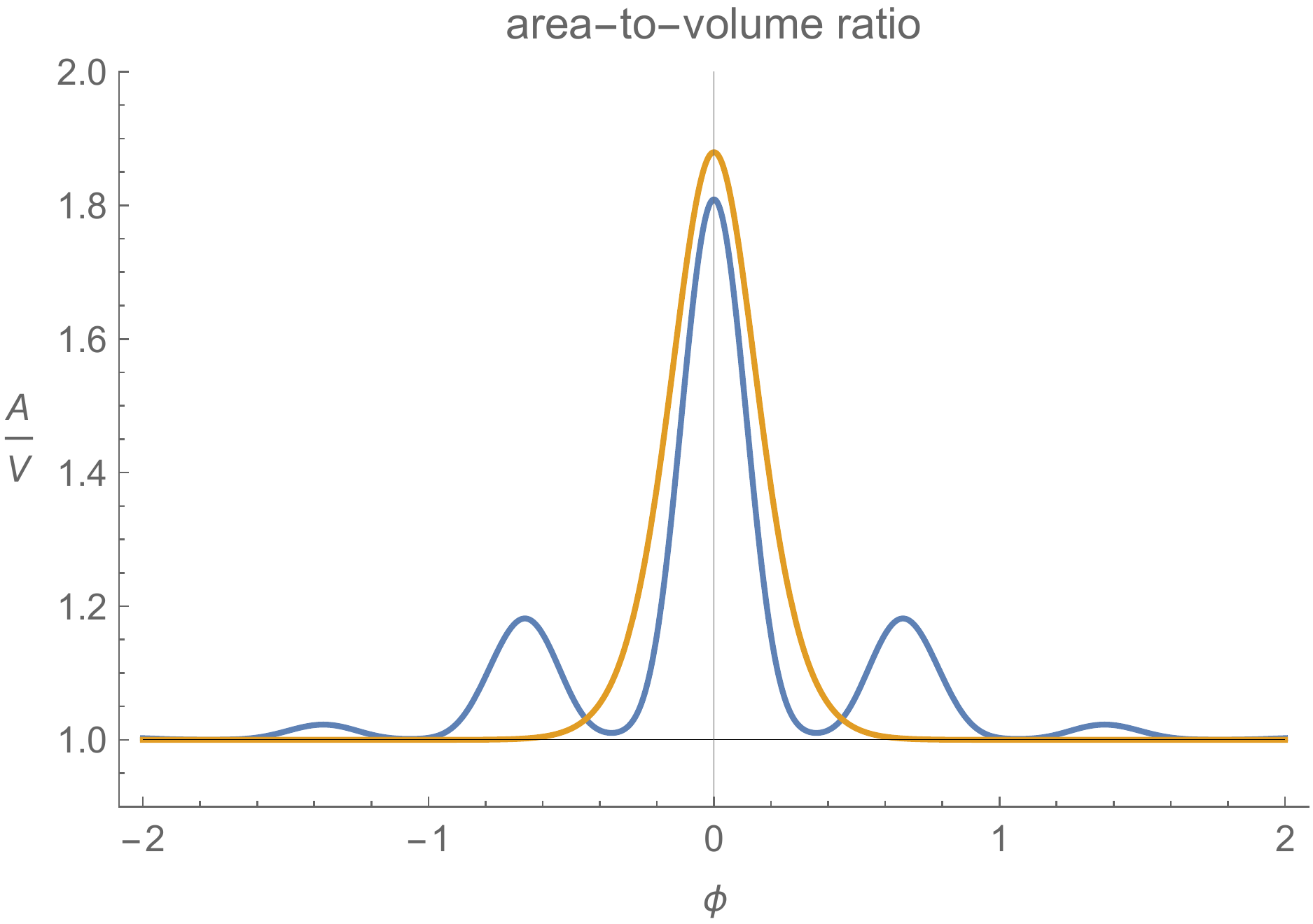}
\caption{Plot of the \emph{surface-area-to-volume ratio} as a function of relational time $\phi$ in the case $\tau<0$, $-6\eta<m^2<-3\eta$. The vertical axis is in units of $\frac{A_0}{V_0}$. The orange curve $(\tau=-1,m^2=-42, Q_0=1, Q_1=1, E_0=-70, E_1=3)$  corresponds to perturbations sitting at the minimum of the potential $U$. Although the initial conditions can be chosen so that the surface-area-to-volume ratio $\frac{A}{V}$ is significantly different from its value for a single tetrahedron at the bounce, it decays exponentially away from it. The blue curve $(\mu=-24, Q_0=3, Q_1=1.5, E_0=2, E_1=14)$ represents the case in which the energy of the perturbations is above the minimum of the potential, but the amplitude of the oscillations is small enough to justify the harmonic approximation. Initial conditions are chosen such that perturbations start oscillating with maximum amplitude at the bounce. The ratio $\frac{A}{V}$ undergoes damped oscillations away from the bounce. The value $\frac{A_0}{V_0}$ is always a lower bound, asymptotically attained. The qualitative behaviour represented by the blue curve is generic for any choice of parameters. When the kinetic energy of the perturbations is negligible compared to the potential energy, one obtains the behaviour represented by the orange curve.}\label{Fig:AtoV}
\end{figure}

The evolution of $\frac{A}{V}$ in case \textit{ii}) is shown in Fig.~\ref{Fig:AtoV}. If the perturbations have minimum ``GFT energy'' (introduced in Eq.~(\ref{eq:DefOmega1}) as $E_1$), \emph{i.e.} they sit at the minimum of $U$, $\frac{A}{V}$ drops off monotonically as we move away from the bounce, due to the growth of the background. One could say that anisotropies, to the extent in which they are captured by the non-monochromatic perturbations, are diluted away by the expansion of the isotropic background. The background value $\frac{A_0}{V_0}$ is a lower bound, which is asymptotically attained in the infinite volume limit (obviously, before too large volumes can be attained, one expects GFT interactions to kick in, breaking the approximation we have employed here). On the other hand, if the ``GFT energy'' of the perturbations is above the minimum of the potential $U$, the perturbations will start to oscillate around such minimum. Therefore, $\frac{A}{V}$ will oscillate as it drops off. The asymptotic properties are unchanged.

\

\begin{figure}
\includegraphics[width=0.6\columnwidth]{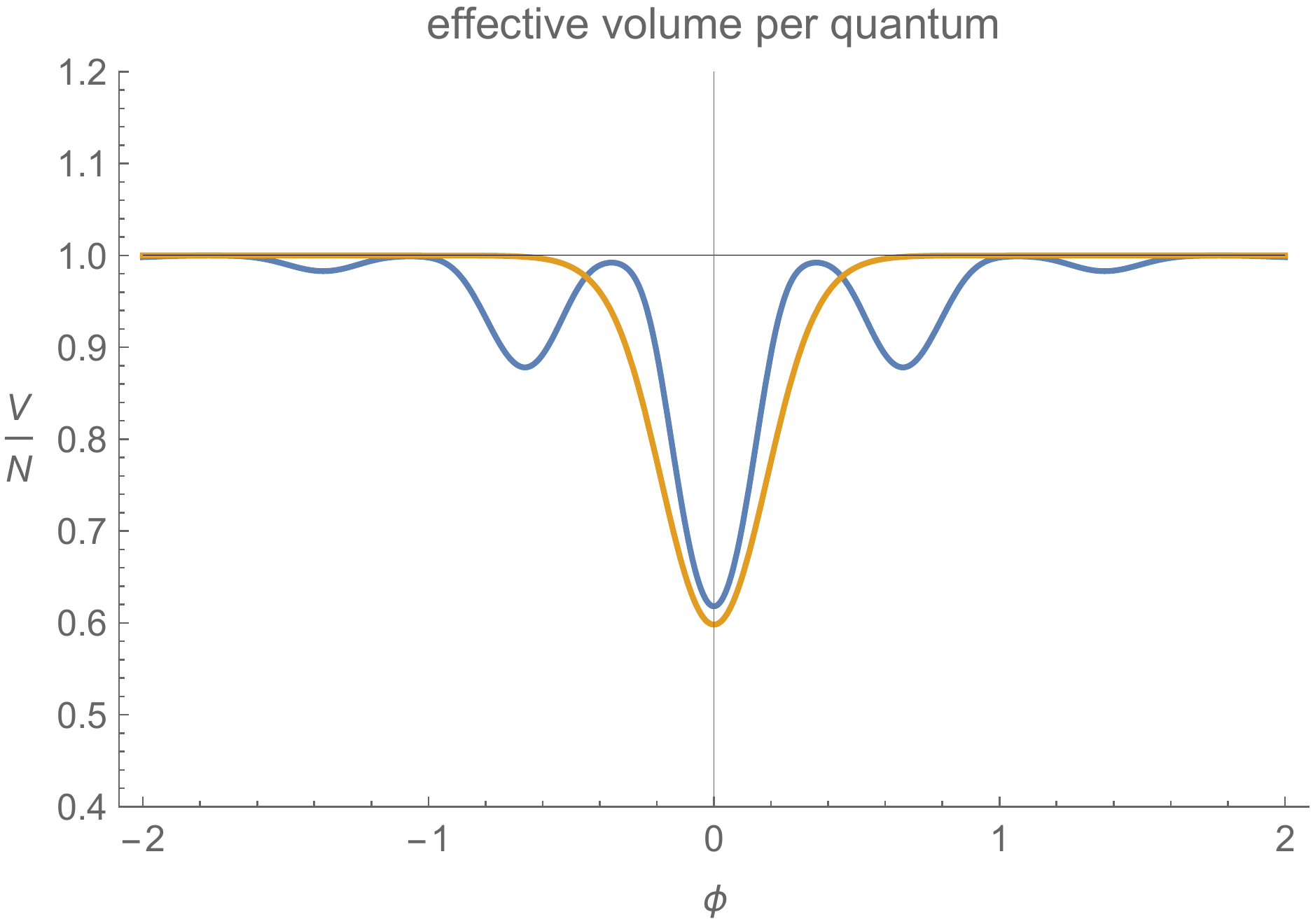}
\caption{Evolution of the effective volume $\frac{V}{N}$ of a quantum over relational time for $\tau<0$, $-6\eta<m^2<-3\eta$. The vertical axis is in units of $V_0$. The parameters chosen for the two curves correspond to those of Fig.~\ref{Fig:AtoV}. $\frac{V}{N}$ relaxes to the volume $V_0$ of a quantum in the backgound away from the bounce. However, at the bounce it can be significantly different from such value.}\label{Fig:EffectiveVolume}
\end{figure}

\begin{figure}
\includegraphics[width=0.6\columnwidth]{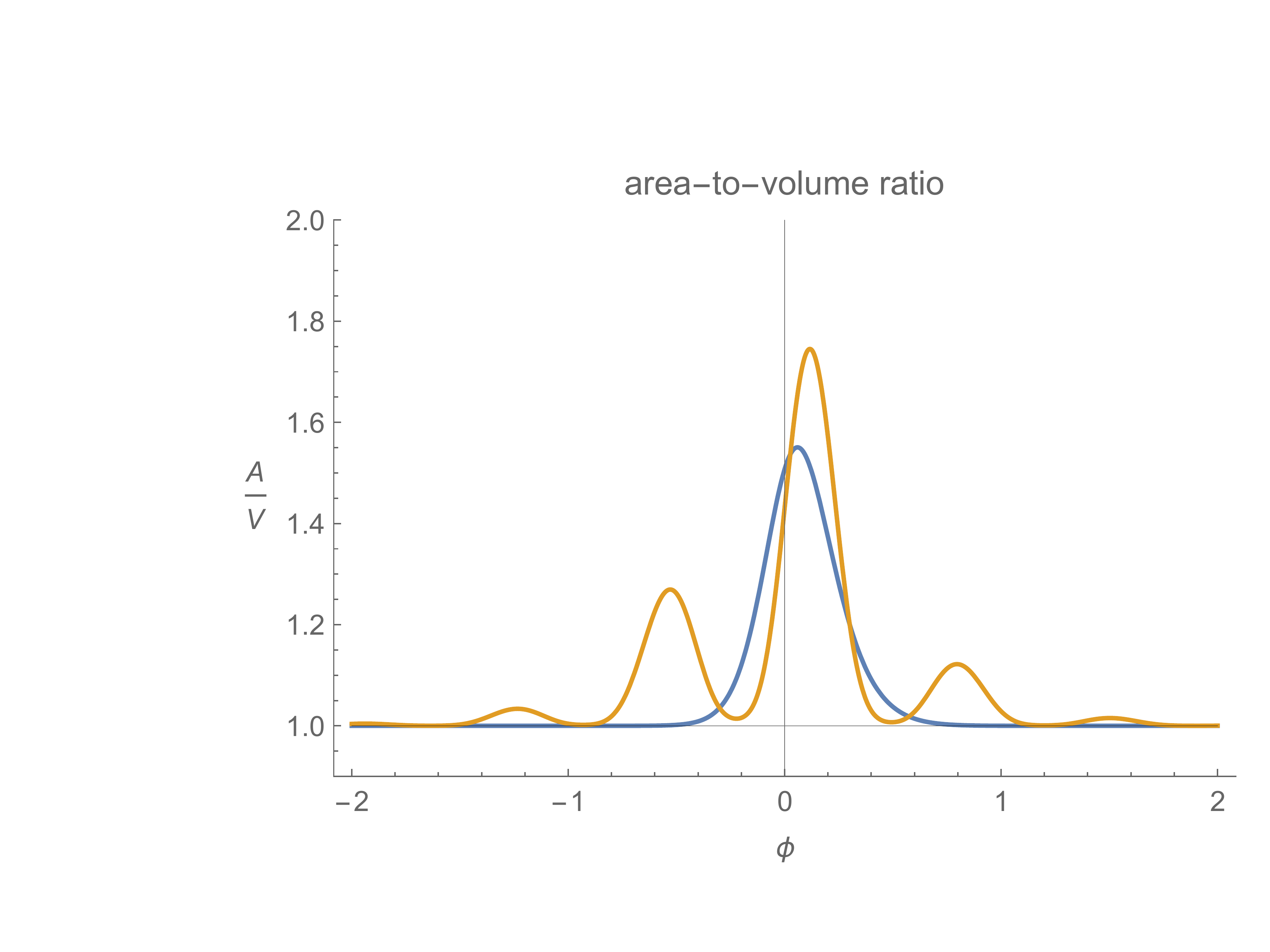}
\caption{Plots of $\frac{A}{V}$ for the case of a non-maximal amplitude of the non-monochromatic perturbations at the bounce. The curves correspond to the same values considered in Fig.~\ref{Fig:AtoV}. An initial ``velocity'' $\pa_{\phi}|\psi|$ is given to the perturbations at the bounce $\phi=\Phi$, with the value 0.6 for the blue curve and 0.4 for the orange curve. The non-symmetric initial conditions results in a deformation of the profile of $\frac{A}{V}$ and is accompanied by a damping.
}\label{Fig:PreBounceIC} 
\end{figure}

Another interesting quantity to compute is the \emph{effective volume per quantum}, defined as
\be
\frac{V}{N}=\frac{ N_0~V_0}{N_0+N_1}=\frac{V_0 }{1+\frac{N_1}{N_0}}\qquad ,
\ee
where again all quantities entering the above formula are expectation values of 2nd quantized GFT observables in the (perturbed) GFT condensate state. 
It satisfies the bounds
\be
0\leq \frac{V}{N}\leq V_0.
\ee
Its profile for the example given above is shown in Fig.~\ref{Fig:EffectiveVolume}. The ratio $\frac{V}{N}$ represents the average volume of a quantum of space. Its value is generally lower than $V_0$, \emph{i.e.} the volume of an equilateral quantum tetrahedron with minimal areas. In fact, zero volume quanta\footnote{See Appendix~\ref{sec:Volume}.} can change the total number of quanta $N$, leaving $V$ unchanged. Explicit calculations show that, in the limit of large $N$, the ratio $\frac{V}{N}$ approaches the value $V_0$ (see Fig.~\ref{Fig:EffectiveVolume}). In Fig.~\ref{Fig:PreBounceIC} we show the plot relative to the case where perturbations do not reach their maximum amplitude at the bounce, resulting in a deformation of the profile of $\frac{A}{V}$. This corresponds to setting initial conditions for the microscopic anisotropies (non-monochromaticity) before the bounce.

\

To summarize, in the region of parameter space corresponding to Case ii) above, our results confirm that, from a bouncing phase, where the quantum geometry can be rather degenerate and  anisotropies (encoded in non-monochromatic perturbations of the simplest GFT condensate state) quite large, a cosmological background emerges whose dynamics can be cast into the form an effective Friedmann equation for a homogeneous, isotropic universe.

\section{Conclusions}
In this article, we have developed further the line of research aimed at understanding cosmology from within full quantum gravity, in the context of the GFT formalism, focusing on the dynamics of anisotropies close to a bouncing region of cosmological evolution. 

In GFT condensate cosmology, the homogeneous cosmological background is identified, in the simplest approximation, with a \emph{field coherent state} in the Fock space of the theory, whose quanta can be pictured as fundamental tetrahedra. To this end, additional conditions interpreted as imposing an isotropic restriction are imposed. It corresponds to a mean field approximation of the complete quantum dynamics of the theory, in which only monochromatic tetrahedra, \emph{i.e.} having identical faces labelled by the same group representation with spin $j$, are excited and all in the same quantum state. A more physical cosmological description in terms of macroscopic geometric quantities is then obtained from the mean field dynamics of the condensate wave function by looking at the evolution of the expectation values of observables (\emph{e.g.} the volume operator) on such coherent states. Applications to cosmology of the quantum gravity model considered in this work were first introduced in Refs.~\cite{\BlockRefJ}, where an effective Friedmann equation was derived that describes the evolution of the background. The most significant consequence of this equation is the resolution of the initial singularity of standard cosmology. Further applications of the model, which are relevant for their implications for the era of accelerated expansion following the bounce, were considered in Refs.~\cite{deCesare:2016axk,deCesare:2016rsf}. In all the mentioned work, the underlying GFT model was chosen to be a generalised version of the EPRL spin foam model for quantum gravity, extensively studied in the loop quantum gravity literature, in which some of the ambiguities in the microscopic dynamics have been suitably parametrised.  

For the same model, we have derived the effective dynamical equations for non-monochromatic perturbations of an isotropic (monochromatic) background, remaining at the mean field level. While a detailed characterization of the anisotropic degrees of freedom in terms of GFT observables with a clear macroscopic, cosmological interpretation is lacking at present, we know that such anisotropic degrees of freedom are in fact encoded in the type of non-monochromatic perturbations we have analysed. Missing a clear geometric interpretation, however, we confined our analysis to a study of the non-monochromatic amplitudes, investigating in which region of parameter space they remain subdominant compared to the isotropic background. Specifically, we derived the classical evolution equations for such perturbations to first order in perturbation theory, in the general case, and then solved them exactly in a simpler case corresponding to a specific choice of EPRL-like GFT model, and for a simple background condensate in which a single spin component is excited and takes a fixed value. We have focused on the approximate regime of these equations corresponding to a cosmic bounce in the evolution of the background, replacing the initial big bang singularity. This is the most interesting regime from the cosmological point of view, where control over anisotropies is critical, but also the regime of GFT mean field dynamics that is technically simpler to study, since in this regime GFT interactions are expected to be subdominant compared to the free dynamics. Then, we determined different regions in the parameter space of the model where perturbations exhibit interesting behaviour; more precisely, we have identified in which region of parameter space, non-monochromatic perturbations decay rapidly away from the bounce, as the universe expands, even if significant close to the bounce. Furthermore, for suitable values of the initial conditions and of the interaction strength, perturbations can become negligible before the interactions kick in. Hence, it is sufficient to consider the monochromatic background in the non-linear regime. Finally, we confirmed the behaviour of such perturbations by a quantitative study of some simple GFT observables: the \emph{surface-area-to-volume ratio} and the effective 1-body volume. Although the relation between such quantities and physical observables with a cosmological interpretation is not clear, they are used in order to illustrate the departures from the case of an isotropic background of monochromatic tetrahedra, previously studied in the literature. 
Our analysis, therefore, strengthens the findings of Refs.~\cite{\BlockRefJ,deCesare:2016axk} that after a bouncing phase, where the quantum geometry can be rather degenerate, a cosmological background emerges whose dynamics can be cast into the form an effective Friedmann equation.

\

It should be clear that our analysis is only a first step towards a more comprehensive study of cosmological anisotropies in the emergent cosmological dynamics of GFT condensates. 

An immediate extension of our work would be to study non-monochromatic perturbations over a different, still isotropic condensate state, and still in a mean field approximation. This would serve the purpose of confirming the general expectation that the ambiguities in associating a continuum geometry to GFT condensate wavefunctions do not affect drastically the effective cosmological dynamics. 

More interestingly, our analysis should be extended with the inclusion of GFT interactions, thus solving the more general equations for the non-monochromatic perturbations we have derived in this paper. This can possibly be done at a phenomenological level, without using the exact analytic expression for the EPRL vertex amplitude. Such an extension would be useful also to obtain a more precise estimate of the regime of validity of the various approximations used in our study, and a more detailed analysis of the dynamics of anisotropies away from the bounce, even if these are confirmed to be subdominant. An interesting question, for example, is what the backreaction of such non-monochromatic perturbations on the background isotropic dynamics is, and what the additional terms in the effective cosmological equations could be that take this backreaction effectively into account. 

More important still, we need to develop a precise characterization of the anisotropic degrees of freedom encoded in non-monochromatic perturbations, in order to be able to describe their dynamics in more explicit, geometric terms. For this, it is necessary to identify suitable observables of clear geometric meaning of measuring cosmological anisotropies, i.e. gauge invariant combinations of the scale factors. In principle, simple condensate states like the ones we have used in this paper are rich enough to capture such observables, at least at the kinematical level, since the domain of definition of the condensate wavefunction is isomorphic to the minisuperspace of generic anisotropic geometries. However, the construction of such suitable observables is far from trivial and has not been carried out so far; it could be that this construction is more naturally carried out by exploiting more involved condensate (or other many-body) states in the GFT Hilbert space, because it may be needed (or at least useful) to rely on the connectivity information present in generic states and absent in the simple coherent states we used. In any case, once a good definition of anisotropic observables is achieved, the effective dynamics of non-monochromatic perturbations should be translated into an effective dynamics for anisotropic geometries, and compared with those expected from classical GR, i.e. Bianchi models. This will be a crucial test of this approach, and at the same time a direction in which it could bring even more interesting fruits.

Needless to say, going beyond the simple mean field approximation of the full quantum GFT dynamics, and to study the corresponding quantum-improved effective cosmological dynamics is an important and interesting task in itself; the same can be said for the inclusion of inhomogeneities in the picture.

\section*{Acknowledgements} 
The authors thank A.~Ashtekar for suggesting to study the surface-area-to-volume ratio to characterize the emergent geometry. We also thank M. Finocchiaro, S. Gielen, E. Wilson-Ewing for discussions and useful comments.

MdC is thankful for the hospitality received at the Max Planck Institute for Gravitational Physics in Potsdam, where part of this work was done. The work of MdC was in part supported by a STSM Grant from COST Action MP1405 QSPACE.

AP gratefully acknowledges the repeated and ongoing hospitality at the MPI for Gravitational Physics. The research of AP was generously supported in part by Perimeter Institute for Theoretical Physics through its Visiting Graduate Fellowship program. Research at Perimeter Institute is supported by the Government of Canada through the Department of Innovation, Science and Economic Development and by the Province of Ontario through the Ministry of Research, Innovation and Science.

\newpage
\appendix

\section{Harmonic Analysis on $\mbox{SU(2)}$}\label{sec:Harmonic}
Harmonic analysis is a generalisation of Fourier analysis to topological groups.
In this appendix we review the fundamentals of harmonic analysis on the Lie group $\mbox{SU(2)}$. We obtain the invariant volume element (Haar measure) and the eigenvectors of the Laplacian operator. The Peter-Weyl formula is then given as a generalisation of the Fourier series expansion.

There is a well known natural omeomorphism between the group manifold $\mbox{SU(2)}$ and the three-sphere $S^3$.
The three dimensional sphere $S^3$ has a natural embedding in Euclidean $\mathbb{R}^4$ as the set of points with Cartesian coordinates $(x_1,x_2,x_3,x_4)$ satisfying the equation
\be\label{eq:constraint}
x_1^2+x_2^2+x_3^2+x_4^2=\eta^{-1}~,
\ee
where $\eta^{-1/2}$ is the radius of the sphere.
A convenient parametrisation of $\mbox{SU(2)}$ elements is given in terms of Euler angles. In fact, every element can be written as
\be
g=\e^{-i\psi \frac{\sigma_3}{2}}\e^{-i\theta \frac{\sigma_2}{2}}\e^{-i\phi \frac{\sigma_3}{2}}=\begin{pmatrix}e^{-\frac{i \phi }{2}-\frac{i \psi }{2}} \cos \left(\frac{\theta }{2}\right) & -e^{\frac{i
   \phi }{2}-\frac{i \psi }{2}} \sin \left(\frac{\theta }{2}\right) \\
 e^{\frac{i \psi }{2}-\frac{i \phi }{2}} \sin \left(\frac{\theta }{2}\right) & e^{\frac{i
   \phi }{2}+\frac{i \psi }{2}} \cos \left(\frac{\theta }{2}\right)\end{pmatrix}=\eta^{1/2}\begin{pmatrix} x_4+ix_3 & ix_1+x_2\\ ix_1-x_2 &x_4-ix_3\end{pmatrix}~.
\ee
The Euler angles have range $0\leq\theta<\pi$, $0\leq\psi<2\pi$, $0\leq\phi<4\pi$.
The metric on $S^3$ is the one induced from the Euclidean metric on $\mathbb{R}^4$
\be\label{eq:MetricEuler}
\de l^2=\frac{\eta^{-1}}{4}\big(\de\theta^2+\de\psi^2+2\cos\theta~\de\psi\,\de\phi+\de\phi^2\big)~.
\ee
The invariant volume element corresponding to the metric (\ref{eq:MetricEuler})
\be\label{eq:VolumeElement}
\de\mu_{\rm Haar}=\frac{\sin\theta}{8}\eta^{-3/2}\;\de\theta\,\de\psi\,\de\phi~.
\ee
This volume element defines the \emph{Haar measure} as the unique measure (up to rescalings) which is invariant under right and left action of the group onto itself.
It is convenient to rescale the volume element by the volume of the group. Integrating Eq.~(\ref{eq:VolumeElement}) over the whole group and normalising the Haar measure to one we have
\be
1=\int_{\rm SU(2)}\de\mu_{\rm Haar}=\frac{1}{8}\eta^{-3/2}\int_0^{2\pi}\de\psi\int_0^{4\pi}\de\phi\int_0^\pi\de\theta\sin\theta=2\pi^2 \eta^{-3/2}~.
\ee
Thus, $\eta$ is determined as
\be\label{eq:EtaValue}
\eta=(2\pi^2)^{2/3}~.
\ee
We define the Laplacian corresponding to the metric in Eq.~(\ref{eq:MetricEuler}) as
\be
\triangle_\eta f=\frac{1}{\sqrt{h}}\,\partial_i\big(\sqrt{h}\,h^{ij}\partial_j f\big)~,
\ee
where $f$ is a smooth function and $h_{ij}$ is the metric on $\mbox{SU(2)}$ given in Eq.~(\ref{eq:MetricEuler}). For $\eta=1$ we have the standard Laplacian on the three-sphere with unit radius, which we denote by $\triangle$.
We observe that
\be
\triangle_\eta =\eta~\triangle .
\ee

Eigenfunctions of the standard Laplacian $\triangle$ are given by the Wigner matrices $D^j_{mn}(\psi,\theta,\phi)$. They satisfy the equation
\be
-\triangle D^j_{mn}=E_j D^j_{mn}~,
\ee
with eigenvalues
\be
E_j=4j(j+1)~.
\ee
The indices $m, n$ corresponding to the same value $j$ are degenerate. Thus, we have for the eigenvalues of the rescaled Laplacian $\triangle_\eta$
\be
E_{j}^{\eta}=\eta~4j(j+1)=\big(2\pi^2\big)^{2/3}\;4j(j+1)~.
\ee

The Wigner matrices form a complete basis of the Hilbert space $L^2(\mbox{SU(2)},\de\mu_{\textrm Haar})$. They satisfy the following orthogonality relations
\be\label{eq:Orthogonality}
\int \de\mu_{\textrm Haar}\; \overline{D^{j_1}_{m_1n_1}}D^{j_2}_{m_2 n_2}=\frac{1}{d_j}\delta_{j_1j_2}\delta_{m_1m_2}\delta_{n_1n_2}~,
\ee
where $d_j=2j+1$ is the dimension of the irreducible representation of $\mbox{SU(2)}$ with spin $j$. Under complex conjugation
\be
\overline{D^j_{mn}}=(-1)^{2j+m+n}D^j_{-m\,-n}~.
\ee

The Peter-Weyl formula gives the decomposition of a square integrable functions on $\mbox{SU(2)}$ in the basis given by the Wigner matrices
\be\label{eq:PeterWeyl}
f(g)=\sum_{j,m,n}f^{j}_{m,n}\sqrt{d_{j}}\;D^{j}_{m,n}(g)~.
\ee
In Section~~\ref{sec:1} we used a straightforward generalisation of Eq.~(\ref{eq:PeterWeyl}) obtained by considering functions defined on the direct product of four copies of $\mbox{SU(2)}$.

\section{Intertwiner Space of a four-valent vertex}\label{sec:Intertwiner}
The Peter-Weyl decomposition of the GFT field can be used to make the degrees of freedom of the theory more transparent. In fact, as observed in Section~\ref{sec:1}, the right-invariance property, Eq.~(\ref{eq:right invariance}), implies that certain coefficients in the series expansion in Eq.~\ref{eq:GFTfield} vanish.
In order to have a non-zero coefficient, the spin labels of the different $\mbox{SU(2)}$ copies must satisfy certain algebraic conditions. In this Appendix, we will first review the constuction of the intertwiner space of a four-valent vertex and then clarify its relation with the kinematics of GFT.

\begin{figure}
\includegraphics[width=0.5\columnwidth]{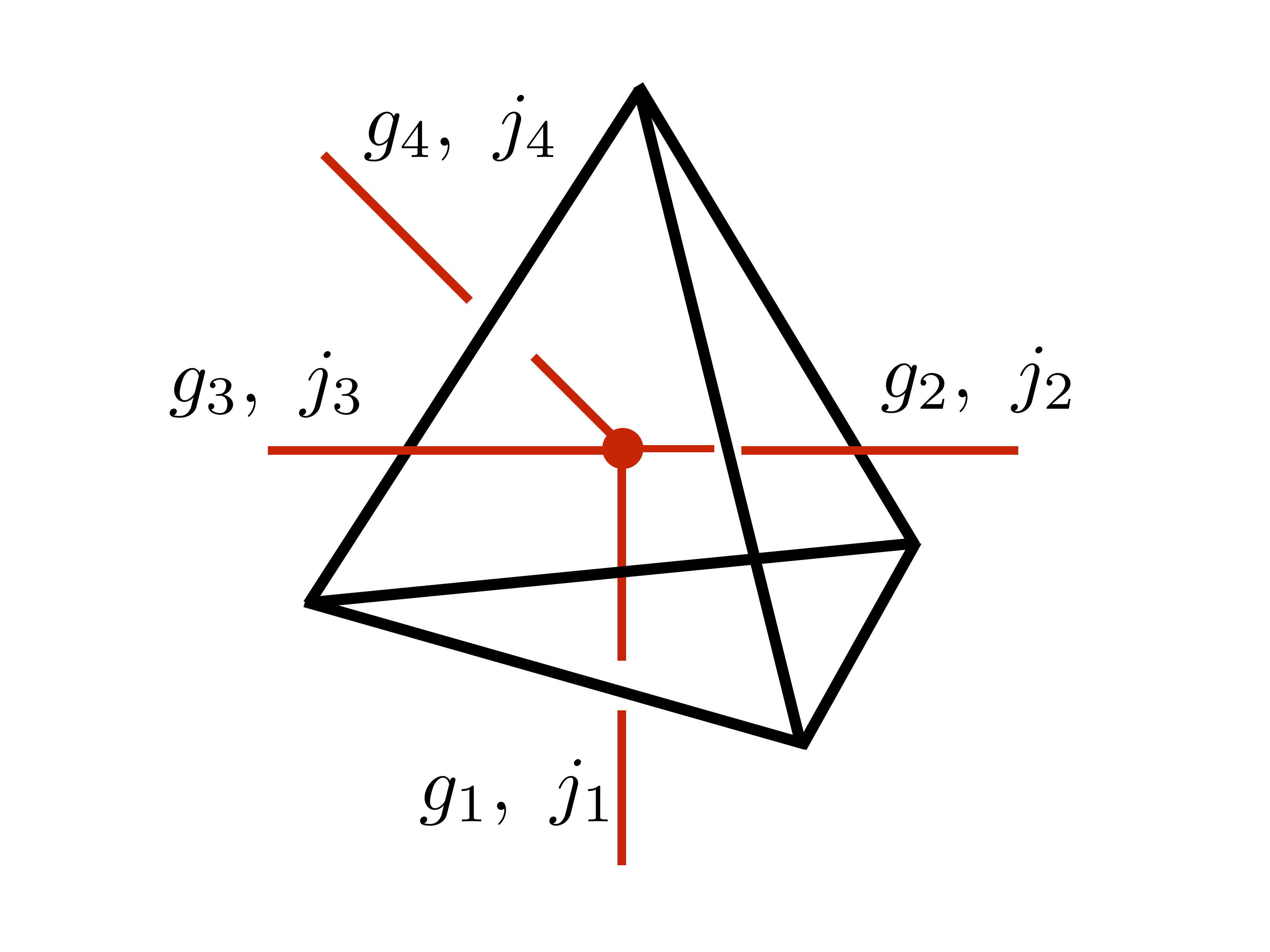}
\caption{An open spin network vertex, corresponding to an elementary excitation of the GFT vacuum. To each link is attached a group element $g_i$ (\emph{holonomy}). The four-valent vertex is dual to a tetrahedron, whose faces are labelled by $\mathfrak{su}(2)$ representations with spin $j_i$. The four spins must satisfy the closure condition Eq.~(\ref{eq:Closure}).}\label{Fig:Tetrahedron}
\end{figure}

In Eq.~(\ref{eq:GFTfield}) the series coefficients are labelled by four spins $j_\nu$. Each of them identifies a finite dimensional Hilbert space $\mathcal{H}_{j_i}$ which is also an irreducible representation of the Lie group $\mbox{SU(2)}$.
The intertwiner space $\accentset{\circ}{\mathcal{H}}_{j_\nu}$ is defined as the subspace of the tensor product $\mathcal{H}_{j_1}\otimes\mathcal{H}_{j_2}\otimes\mathcal{H}_{j_4}\otimes\mathcal{H}_{j_4}$ whose elements are invariant under the diagonal action of $\mbox{SU(2)}$, \emph{i.e.} we define it as the space of invariant tensors \cite{Haggard:2011qvx}
\be
\accentset{\circ}{\mathcal{H}}_{j_\nu}=\mbox{Inv}_{\mbox{\scriptsize SU(2)}}\left[\mathcal{H}_{j_1}\otimes\mathcal{H}_{j_2}\otimes\mathcal{H}_{j_3}\otimes\mathcal{H}_{j_4}\right]~.
\ee
Thus, $\accentset{\circ}{\mathcal{H}}_{j_\nu}$ is the space of singlets that can be constructed out of four spins. It can be interpreted as the Hilbert space of a quantum tetrahedron. In fact, we can give a geometric interpretation of the construction of the intertwiner space. Let us consider four links, each carrying a spin $j_i$ and meeting at a vertex. A tetrahedron is constructed by duality from the vertex; the spins $j_i$ are the quantum numbers of the areas of its faces. Global invariance of the vertex under $\mbox{SU(2)}$ amounts to the closure of the tetrahedron, see Fig.~\ref{Fig:Tetrahedron}.

A basis in $\accentset{\circ}{\mathcal{H}}_{j_\nu}$ can be found by first composing the four spins pairwise, then the two resultant spins together so as to form singlets. We partition the spins in two pairs $(j_1,j_2)$ and $(j_3,j_4)$, corresponding to the recoupling channel $\mathcal{H}_{j_1}\otimes\mathcal{H}_{j_2}$ \cite{Haggard:2011qvx}. The total spin of a pair is labelled by the quantum number $J$, which is the same for each of the two pairs since they sum to give a singlet. Basis vectors can thus be expressed in terms of the tensor product basis as
\be\label{eq:Singlets}
|j_\nu ;J\rangle=\sum_{m_{\nu}}\alpha^{j_\nu,J}_{m_\nu}|j_1,m_1\rangle|j_2,m_3\rangle|j_3,m_3\rangle|j_4,m_4\rangle~.
\ee
The coefficients $\alpha^{j_\nu,J}_{m_\nu}$ are the elements of a unitary matrix, which implements the change of basis from the tensor product basis $\{|j_1,m_1\rangle|j_2,m_3\rangle|j_3,m_3\rangle|j_4,m_4\rangle\}$ to $\{|j_\nu;J\rangle\}$ in the space of singlets $\accentset{\circ}{\mathcal{H}}_{j_\nu}$~\footnote{Notice that it is not a unitary matrix over the whole Hilbert space $\mathcal{H}_{j_1}\otimes\mathcal{H}_{j_2}\otimes\mathcal{H}_{j_4}\otimes\mathcal{H}_{j_4}$, since $\alpha^{j_\nu,J}_{m_\nu}$ vanishes when $m_\nu$ fails to satisfy Eq.~(\ref{eq:SpinsSumToZero})}.
$\alpha^{j_\nu,J}_{m_\nu}$ is an invariant tensor, \emph{i.e.} all of its components are invariant under $\mbox{SU(2)}$.

The quantum number $J$ satisfies the inequalities
\be\label{eq:InequalityJ}
\mbox{max}\left\{|j_1-j_2|,|j_3-j_4|\right\}\leq J\leq \mbox{min}\left\{j_1+j_2,j_3+j_4\right\}.
\ee
Moreover, in order to get a singlet one must have
\be\label{eq:SpinsSumToZero}
m_1+m_2+m_3+m_4=0~.
\ee
When Eqs.~(\ref{eq:InequalityJ}),~(\ref{eq:SpinsSumToZero}) are not satisfied for certain values of $J$ and $m_\nu$, $\alpha^{j_\nu,J}_{m_\nu}$ vanishes and the corresponding term gives a vanishing contribution to Eq.~(\ref{eq:Singlets}).
The coefficients of the decomposition in Eq.~(\ref{eq:Singlets}) can be expressed in terms of Clebsch-Gordan coefficients as
\be\label{eq:SingletSpinJ}
\alpha^{j_\nu,J}_{m_\nu}=\eta~\frac{(-1)^{J-M}}{\sqrt{d_J}}C^{j_1j_2J}_{m_1m_2\,M}C^{j_3j_4J}_{m_3m_4\,-M}~,
\ee
where we defined $M=m_1+m_2=-(m_3+m_4)$ and $\eta$ is a phase factor. The latter can depend on $J$ as well as on the fixed values of the four spins $j_\nu$. The functional dependence is omitted to avoid confusion with tensor indices. Clearly, the value of $\eta$ does not affect the unitarity relation satisfied by the coefficients defined in Eq.~(\ref{eq:SingletSpinJ})
\be
\sum_{m_\nu}\overline{\alpha}^{j_\nu,J}_{m_\nu}\alpha^{j_\nu,J^{\prime}}_{m_\nu}=\delta^{J J^{\prime}}~.
\ee
We choose the value of the phase $\eta$ so as to have
\be\label{eq:DefAlphaIntertwiner}
\alpha^{j_\nu,J}_{m_\nu}=(-1)^{J-M}\sqrt{d_J}\left(\begin{array}{ccc} j_1 & j_2 & J\\ m_1 & m_2 & -M\end{array}\right)\left(\begin{array}{ccc} J & j_4 & j_3\\ M & m_4 & m_3\end{array}\right)~.
\ee
The convenience of this particular choice of conventions lies in the fact that the contraction of five four valent intertwiners thus defined coincides with the definition of the 15j symbol, \emph{i.e.}

All intertwiners, \emph{i.e.} elements of $\accentset{\circ}{\mathcal{H}}_{j_\nu}$ can be expressed as linear combinations of the $\alpha^{j_\nu,J}_{n_\nu}$ coefficients given above
\be\label{eq:LinearCombinationIotaAlpha}
\mathcal{I}^{j_\nu,\iota}_{n_\nu}=\sum_{J} c^{J\iota}\,\alpha^{j_\nu,J}_{n_\nu}~.
\ee
 Hence,  we can attach the label $\iota$ to any linear subspace in the intertwiner space $\accentset{\circ}{\mathcal{H}}_{j_\nu}$. Different choices correspond to different physical properties of the `quanta' of geometry.
  In Appendix~\ref{sec:Volume} we construct explicitly the intertwiners of volume eigenstates in a simple example.

\section{Volume Operator}\label{sec:Volume}
There are several different definitions of the volume operator in LQG \cite{Ashtekar:1997fb},~\cite{Rovelli:1994ge},~\cite{Bianchi:2010gc}. However, they all agree in the case of a four-valent vertex \cite{Haggard:2011qvx} and match the operator introduced in Ref.~\cite{Barbieri:1997ks}. In this Appendix we will largely follow Ref.~\cite{Brunnemann:aa} for the definition of the volume operator and the derivation of its spectrum. The volume operator acting on a spin network vertex (embedded in a differentiable manifold) is defined as
\be\label{eq:VolumeDefinition}
\hat{V}=\sqrt{\left|\sum_{I<J<K}\epsilon(e_I,e_J,e_K)\epsilon_{ijk}J^i_I J^j_J J^k_K\right|}=\sqrt{\left|\frac{i}{4}\sum_{I<J<K}\epsilon(e_I,e_J,e_K)\hat{q}_{IJK}\right|}~.
\ee
In the formula above $(e_I,e_J,e_K)$ is a triple of edges adjacent to the vertex and $\epsilon(e_I,e_J,e_K)$ is their orientation, given by the triple product of the vectors tangent to the edges. A copy of the angular momentum algebra is attached to each edge, \emph{i.e.} there is one spin degree of freedom per edge. Angular momentum operators corresponding to distinct edges commute
\be
\left[J^i_I,J^j_J\right]=i \delta_{IJ}\epsilon^{ijk}J^k_I~.
\ee
In the last step of Eq.~(\ref{eq:VolumeDefinition}) we introduced the operator
\be
\hat{q}_{IJK}=\left(\frac{2}{i}\right)^3\epsilon_{ijk}J^i_I J^j_J J^j_K~.
\ee

Spin network vertices are gauge-invariant, \emph{i.e.} the angular momenta carried by the edges entering a vertex satisfy a closure condition. In the case of a four-valent vertex the closure condition reads as
\be\label{eq:Closure}
\mathbf{J}_1+\mathbf{J}_2+\mathbf{J}_3+\mathbf{J}_4=\mathbf{0}~.
\ee
Hence, the Hilbert space of the vertex is that of Eq.~(\ref{eq:HilbertSpace4Vertex}).
Eq.~(\ref{eq:Closure}) leads to the following simplification in the evaluation of the sum in Eq.~(\ref{eq:VolumeDefinition})
\be
\sum_{I<J<K}\epsilon(e_I,e_J,e_K)\hat{q}_{IJK}=2\,\hat{q}_{123}~.
\ee
Therefore, the squared volume operator can be rewritten as
\be\label{eq:VolumeExpressionQ}
\hat{V}^2=\left|\frac{i}{2}\hat{q}_{123}\right|~.
\ee
Note that, while the definition (\ref{eq:VolumeDefinition}) makes explicit reference to the embedding map, the final expression (\ref{eq:VolumeExpressionQ}) clearly does not depend on it.

Using the recoupling channel $\mathcal{H}_{j_1}\otimes\mathcal{H}_{j_2}$ as in Appendix~\ref{sec:Intertwiner} and labelling with $J$ the eigenvalue of $(\mathbf{J}_1+\mathbf{J}_2)^2$, we find that the non-vanishing matrix elements in the recoupling basis are \cite{Brunnemann:aa}
\be\label{eq:VolumeSpectrum}
\begin{split}
\langle J | \hat{q}_{123} |J-1\rangle&=\frac{1}{\sqrt{4J^2-1}}\Big[(j_1+j_2+J+1)(-j_1+j_2+J)(j_1-j_2+J)(j_1+j_2-J+1)\Big.\\
&\phantom{=\frac{1}{\sqrt{4J^2-1}} [}  \Big.(j_3+j_4+J+1)(-j_3+j_4+J)(j_3-j_4+J)(j_3+j_4-J+1)\Big]^{\frac{1}{2}}\\ &=-\langle J -1 | \hat{q}_{123} |J\rangle~.
\end{split}
\ee
The eigenvalues of $\hat{q}_{123}$ are non-degenerate. Moreover, if $\hat{q}_{123}$ has a non-vanishing eigenvalue $a$, also $-a$ is an eigenvalue. The sign corresponds to the orientation of the vertex.
If the dimension of the intertwiner space is odd, $\hat{q}_{123}$ has a non-degenerate zero eigenvalue.
\subsection*{Monochromatic Vertex}
If the four spins are all identical ($j_1=j_2=j_3=j_4=j$) the vertex is called monochromatic. In this case Eq.~(\ref{eq:VolumeSpectrum}) simplifies to
\be\label{eq:SpectrumMonochromatic}
\langle J | \hat{q}_{123} |J-1\rangle=\frac{1}{\sqrt{4J^2 -1}} J^2 (d_j^2 - J^2 )~,
\ee
where $d_j=2j+1$ is the dimension of the irreducible representation with spin $j$ \cite{Brunnemann:aa}.

For the applications of Sections~\ref{sec:Friedmann},~\ref{sec:Perturbations}, it is particularly interesting to consider the fundamental representation $j=\frac{1}{2}$. In this case the intertwiner space is two-dimensional, with a basis given by $\{|0\rangle,|1\rangle\}$, \emph{i.e.} the four-valent gauge-invariant vertex is constructed using two singlets and two triplets, respectively. Using Eq.~(\ref{eq:SpectrumMonochromatic}), we have that the squared volume operator is written in this basis as
\be
\hat{V}^2=\frac{\sqrt{3}}{2}\left|\begin{pmatrix}0 & -i \\ i & 0 \end{pmatrix}\right|=\left|\hat{Q}\right|~,
\ee
where we introduced a new matrix $\hat{Q}=\frac{\sqrt{3}}{2}\sigma_2$, which is equal to $\hat{V}^2$ up to a sign. The sign of the eigenvalues of $\hat{Q}$ gives the orientation of the vertex.

The normalised eigenvectors of $\hat{Q}$ are
\be
|+\rangle=\frac{1}{\sqrt{2}}\begin{pmatrix}1\\i\end{pmatrix}, \hspace{1em} |-\rangle=\frac{1}{\sqrt{2}}\begin{pmatrix}1\\-i\end{pmatrix}~,
\ee
with eigenvalues $\pm\frac{\sqrt{3}}{2}$. The volume eigenstates $|\pm\rangle$ can be decomposed in the tensor product basis of $\mathcal{H}_{\frac{1}{2}}\otimes\mathcal{H}_{\frac{1}{2}}\otimes\mathcal{H}_{\frac{1}{2}}\otimes\mathcal{H}_{\frac{1}{2}}$ as follows
\be
|\pm\rangle=\sum_J c^{J\,\pm} |J\rangle=\sum_{m_{\nu}J}c^{J\,\pm}\alpha^{\frac{1}{2}\,J}_{m_\nu}|{\scriptstyle\frac{1}{2}},m_1\rangle|{\scriptstyle\frac{1}{2}},m_2\rangle|{\scriptstyle\frac{1}{2}},m_3\rangle|{\scriptstyle\frac{1}{2}},m_4\rangle~.
\ee
We define the intertwiners corresponding to the volume eigenstates $|\pm\rangle$ as
\be
\mathcal{I}^{\frac{1}{2}\,\pm}_{m_\nu}=\sum_{J}c^{J \pm}\alpha^{\frac{1}{2}\,J}_{m_\nu}=\frac{1}{\sqrt{2}}\left(\alpha^{\frac{1}{2}\,0}_{m_\nu}\pm i\, \alpha^{\frac{1}{2}\,1}_{m_\nu}\right)~.
\ee
Hence, we can write
\be
|\pm\rangle=\sum_{m_{\nu}}\mathcal{I}^{\frac{1}{2}\,\pm}_{m_\nu}|{\scriptstyle\frac{1}{2}},m_1\rangle|{\scriptstyle\frac{1}{2}},m_2\rangle|{\scriptstyle\frac{1}{2}},m_3\rangle|{\scriptstyle\frac{1}{2}},m_4\rangle~.
\ee
%
%

\bibliography{referencesBdG}

\end{document}